\documentclass{emulateapj}
\usepackage{natbib}

\slugcomment{Submitted to ApJ}

\shorttitle{X-ray Morphological Properties of SNRs}
\shortauthors{LOPEZ ET AL.}

\newcommand{\ltsima}{$\; \buildrel < \over \sim \;$}
\newcommand{\simlt}{\lower.5ex\hbox{\ltsima}}

\newcommand{\ls}{{_<\atop^{\sim}}}

\newcommand{\gs}{{_>\atop^{\sim}}}

\begin{document}

\title{Using the X-ray Morphology of Young Supernova Remnants to Constrain Explosion Type, Ejecta Distribution, and Chemical Mixing}

\author{
Laura A. Lopez\altaffilmark{1}, Enrico Ramirez-Ruiz\altaffilmark{1}, Daniela Huppenkothen\altaffilmark{2}. Carles Badenes\altaffilmark{3,4}, David A. Pooley\altaffilmark{5}
}
\altaffiltext{1}{Department of Astronomy and Astrophysics, University of California Santa Cruz, 159 Interdisciplinary Sciences Building, 1156 High Street, Santa Cruz, CA 95064, USA; lopez@astro.ucsc.edu.}
\altaffiltext{2}{Astronomical Institute, Anton Pannekoek, University of Amsterdam, P.O. Box 94249, 1090 GE Amsterdam, Netherlands}
\altaffiltext{3}{School of Physics and Astronomy, Tel-Aviv University, Tel-Aviv 59978, Israel}
\altaffiltext{4}{Benoziyo Center for Astrophysics, Weizmann Institute of Science, Rehovot 76100, Israel}
\altaffiltext{5}{Eureka Scientific, Inc., Austin, TX 78756}

\begin{abstract}

Supernova remnants (SNRs) are a complex class of sources, and their heterogeneous nature has hindered the characterization of their general observational properties. To overcome this challenge, in this paper, we use statistical tools to analyze the {\it Chandra} X-ray images of Galactic and Large Magellanic Cloud SNRs. We apply two techniques, a power-ratio method (a multipole expansion) and wavelet-transform analysis, to measure the global and local morphological properties of the \hbox{X-ray line} and thermal emission in twenty-four SNRs. We find that Type Ia SNRs have statistically more spherical and mirror symmetric thermal X-ray emission than core-collapse (CC) SNRs. The ability to type SNRs based on thermal emission morphology alone enables, for the first time, the typing of SNRs with weak X-ray lines and those with low resolution spectra. Based on our analyses, we identify one source (SNR G344.7$-$0.1) as originating from a CC explosion that was of unknown origin previously; we also confirm the tentative Type Ia classifications of G337.2$-$0.7 and G272.2$-$3.2. Although the global morphology is indicative of the explosion type, the relative morphology of the X-ray line emission within SNRs is not: all sources in our sample have well-mixed ejecta, irrespective of stellar origin. In particular, we find that 90\% of the bright metal-line emitting substructures are spatially coincident and have similar scales, even if the metals arise from different burning processes. Moreover, the overall X-ray line morphologies within each SNR are the same, with $<$6\% differences. These findings reinforce observationally that hydrodynamical instabilities can efficiently mix ejecta in Type Ia and CC SNRs. The only exception is W49B, which can be attributed to its jet-driven/bipolar explosive origin. Based on comparative analyses across our sample, we describe several observational constraints that can be used to test hydrodynamical models of SNR evolution; notably, the filling factor of X-ray emission decreases with SNR age. 

\end{abstract}


\section{Introduction}

Supernova remnants (SNRs) are a diverse class of objects that play an essential role in the Universe, including driving the dynamics of the interstellar medium (ISM) and producing and distributing most of the metals \citep{fp04}. The morphology and dynamics of young SNRs depend on the distribution of the ambient medium and on the structure of the stellar ejecta. Self-similar, spherically-symmetric solutions exist \citep{chev82}, and they are used widely to interpret observational data of young SNRs. However, the ejecta are subject to hydrodynamical instabilities that preclude a self-similar description of the expansion, and thus the use of hydrodynamical models is necessary. 

A major difficulty at present is bridging these hydrodynamical models with observations of young SNRs. A few direct observables of individual sources (e.g., expansion rates) can be compared easily to theoretical predictions. However, the complexity and heterogeneous nature of SNRs has limited the ability to define the observed properties of SNRs as a class. As a consequence, previous observational SNR work has largely focused on interpretation of single objects, without systematic comparison between sources (with some exceptions: e.g., Badenes et al. 2010; Long et al. 2010). Although each SNR is unique and complicated when studied in detail, it is vital to unify the observed characteristics of SNRs to test and to improve hydrodynamical models of their dynamics.

With the advent of high-resolution, space-based telescopes in the last couple decades, the time is ripe to undertake this task. In particular, the {\it Chandra} X-ray Observatory has facilitated an unprecedented view of young ejecta-dominated SNRs since its launch in 1999. The sub-arcsecond spatial resolution and the spatially-resolved spectroscopy capabilities of {\it Chandra} have facilitated detailed studies of the metal-rich ejecta from SN explosions as well as their interactions with the surrounding as they expand (see reviews by Weisskopf \& Hughes 2006; Badenes 2010). {\it Chandra} has observed over one-hundred SNRs in the Milky Way galaxy \citep{green} and many others in nearby galaxies (e.g., M33: Long et al. 2010). This wealth of data provides the necessary basis to characterize the observed X-ray properties of SNRs as a class. 

\begin{deluxetable*}{ccccccccc} \label{sources}
\tablecolumns{9} \tablewidth{0pc} \tabletypesize{\footnotesize}
\setlength{\tabcolsep}{0.0in} \renewcommand{\arraystretch}{1.0}
\tablewidth{0pt}
\tablecaption{Sources, Sorted by Age}
\footnotesize
\tablehead{\colhead{Number} & \colhead{Source} & \colhead{ObsID} & \colhead{ACIS Exp.} & \colhead{Age\tablenotemark{a}} & \colhead{Distance} & \colhead{Radius\tablenotemark{b}} & \colhead{$L_{\rm X}$\tablenotemark{c}} & \colhead{References} \\ 
\colhead{} & \colhead{} & \colhead{} & \colhead{(ks)} & \colhead{(years)} & \colhead{(kpc)} & \colhead{(pc)} & \colhead{($\times 10^{37}$ erg s$^{-1}$)} & \colhead{}}
\startdata
\cutinhead{Type Ia Sources} 
1 & 0509$-$67.5 & 776, 7635, 8554 & 113 & 350--450 & 50 & 5.96 & 1.76 & 1 \\
2 & Kepler & 6714--6718, 7366 & 751 & 405 & 5.0 & 3.88 & 0.19 & 2 \\
3 & Tycho & 3887 &  150 & 437 & 2.4 & 3.72 & 0.12 & 3 \\
4 & 0519$-$69.0 & 118 & 40 & 400--800 & 50 & 6.56 & 1.06 & 4 \\
5 & N103B & 125 & 37 & 860 & 50 & 5.96 & 1.67 & 5 \\
6 & G337.2$-$0.7 & 2763 & 49 & $\sim$750--10000 & 8.0 & 7.63 & 0.71 & 6 \\
7 & DEM L71 & 775, 3876, 4440 & 148 & $\sim$4360 & 50 & 11.9 & 0.77 & 7 \\ 
8 & 0548$-$70.4 & 1992 & 60 & $\sim$7100 & 50 & 17.9 & 0.96 & 8 \\
9 & G272.2$-$3.2 & 9147, 10572 & 65 & $\sim$6250--15250 & 5.0 & 13.12 & 0.05 & 9 \\
10 & 0534$-$69.9 & 1991 & 60 & $\sim$10000 & 50 & 21.47 & 0.97 & 8 \\
\cutinhead{Core-collapse Sources}
11 & Cas A & 4634--4639, 5196, 5319--5320 & 993 & 309--347 & 3.4 & 3.77 & 2.58 & 10 \\
12 & W49B & 117 & 55 & $\sim$1000 & 8.0 & 6.30 & 4.48 & 11 \\
13 & G15.9$+$0.2 & 5530, 6288, 6289 & 30 & $\sim$1000 & 8.5 & 8.72 & 1.38 & 12 \\
14 & G11.2$-$0.3 & 780--781, 2322, 3909--3912 & 95 & 1623 & 5.0 & 4.17 & 1.18 & 13 \\
15 & Kes 73 & 729 & 30 & 500--2200 & 8.0 & 6.11 & 0.94 & 14 \\
16 & RCW 103 & 970 & 49 & $\sim$2000 & 3.3 & 5.43 & 2.21 & 15 \\
17 & N132D & 5532, 7259, 7266 & 90 & $\sim$3150 & 50 & 21.47 & 9.92 & 16 \\
18 & G292.0$+$1.8 & 6677--6680, 8221, 8447 & 516 & $\sim$3300 & 6.0 & 9.45 & 0.56 & 17 \\ 
19 & 0506$-$68.0 & 2762 & 38 & $\sim$4600  & 50 & 17.89 & 2.62 & 18 \\
20 & Kes 79 & 1982 & 30 & $\sim$6400 & 7.1 & 13.55 & 0.10 & 19 \\
21 & N49B & 1041 & 35 & 10000 & 50 & 23.85 & 2.39 & 20 \\
22 & B0453$-$685 & 1990 & 40 & 13000 & 50 & 20.27 & 0.54 & 21 \\
23 & N206 & 3848, 4421 & 69 & $\sim$25000 & 50 & 29.82 & 0.13 & 22 \\
\cutinhead{Unknown Type}
24 & G344.7$-$0.1 & 4651, 5336 & 27 & -- & 14.0 & 18.37 & 4.23 & 23 \\
\enddata
\tablenotetext{a}{Aside from historical SNRs and objects with detected light echoes, these ages are very uncertain.}
\tablenotetext{b}{Radius $R$, selected to enclose the entire source in the full-band X-ray image, and it is determined assuming the distances listed above.}  
\tablenotetext{c}{X-ray luminosity in the 0.3--2.1 keV band, from the {\it Chandra} SNR Catalog. The values for G15.9$+$0.2, G272.2$-$3.2, and G344.7$-$0.1 is calculated using an absorbed planar shock (vpshock) model of the integrated spectra \citep{reynolds06}.}
\tablerefs{[1]~\cite{bad08}; [2]~\cite{reynolds}; [3]~\cite{warren05}; [4]~\cite{rest05}; [5]~\cite{lewis03}; [6]~\cite{rakowski};  [7]~\cite{hughes03}; [8]~\cite{hendrick}; [9]~\cite{harrus}; [10]~\cite{hwang04}; [11]~\cite{lal}; [12]~\cite{reynolds06}; [13]~\cite{kaspi01}; [14]~\cite{gotthelf}; [15]~\cite{carter}; [16]~\cite{bork07}; [17]~\cite{park07}; [18]~\cite{hughes06}; [19]~\cite{sun};  [20]~\cite{park03}; [21]~\cite{gaensler03}; [22]~\cite{williams05}; [23]~\cite{yama}}
\end{deluxetable*}

Toward this end, in this paper, we use quantitative methods to examine the {\it Chandra} images of all SNRs with strong X-ray line and thermal emission in the Milky Way and Large Magellanic Cloud (LMC). We apply well-established mathematical tools to characterize the global and local morphological properties of SNRs; the statistical approach we take here enables, for the first time, the capability to compare our results within and between sources to set observational constraints on hydrodynamical models. 

This paper is organized as follows. In $\S$2, we describe the sample and observations used in this study, and in $\S$3, we present the methods employed in our analyses. In $\S$4, we give our results, and in $\S$5, we examine how morphological properties vary across our sample. $\S$6 discusses the observational constraints on hydrodynamical models we have found in this paper and the implications of our findings.

\section{Observations and Data Preparation}

\begin{figure*}
\epsscale{1.0}
\includegraphics[width=1.0\textwidth]{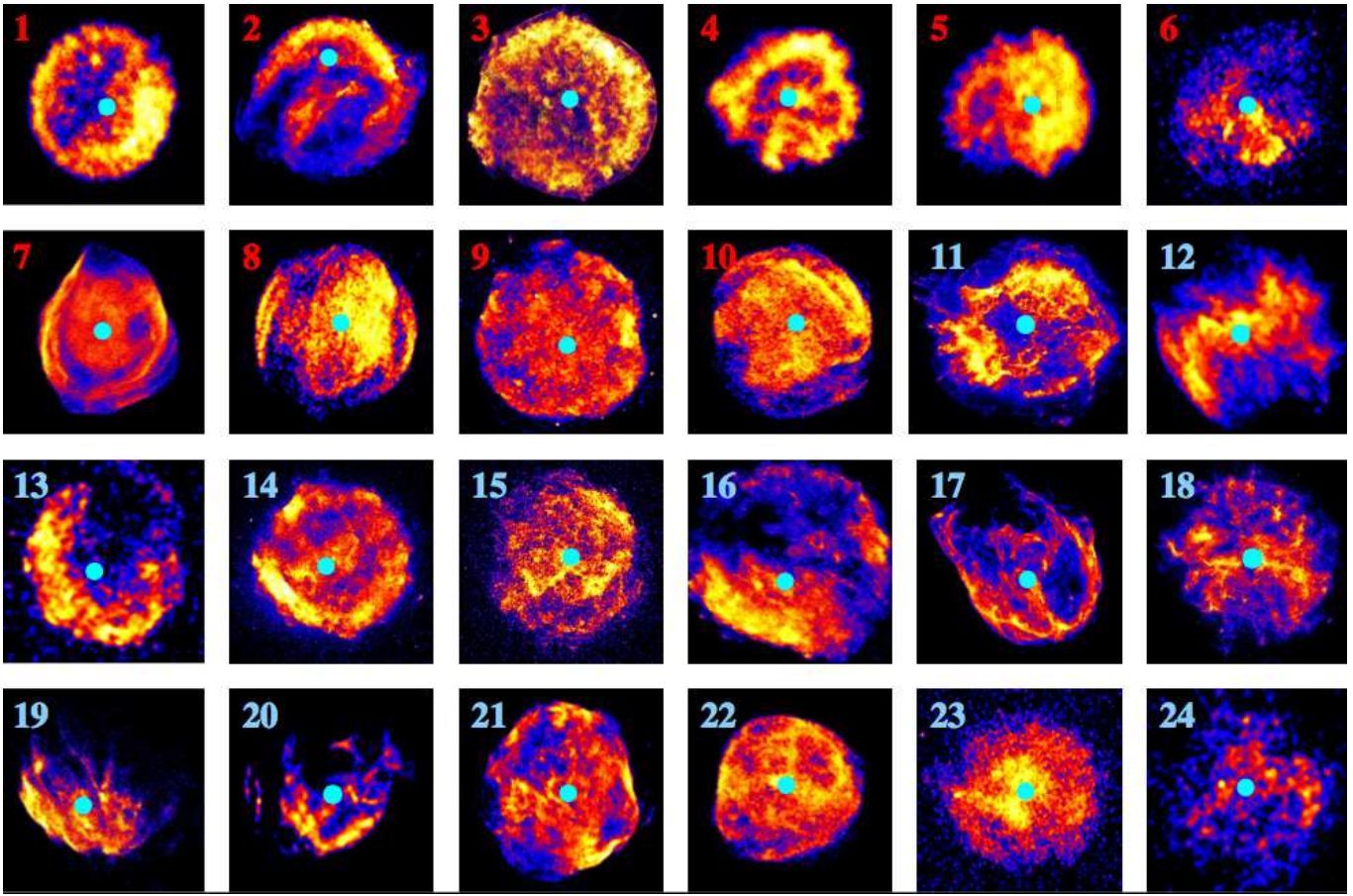}
\caption{{\it Chandra} X-ray soft-band (0.5--2.1 keV) images of the 24 SNRs listed in Table 1. The cyan circles mark the full-band centroids of each SNR used in our power-ratio/multipole expansion method. Numbers correspond to those in Column 1 of Table 1. Red numbers denote Type Ia SNRs; light blue numbers denote CC SNRs.}
\label{fig:stamps}
\end{figure*}

For our analyses, we utilize archival {\it Chandra} ACIS observations of the twenty-four SNRs listed in Table 1 and shown in Figure~\ref{fig:stamps}. Ten of our sources are thought to be from Type Ia SNe, thirteen are considered to have originated from CC SNe, and one is unknown (see Table 1). We selected sources in the Milky Way and LMC that have prominent thermal emission from ejecta in their global X-ray spectra (with 0.5--2.1 keV X-ray counts per unit area $>$0.01 counts/pixel$^{2}$ within the radius $R$ that encloses their signal in Table 1). This criterion excludes SNRs dominated by non-thermal emission, like G1.9$+$0.3 \citep{g1.9}. We also required that the sources be imaged fully in one ACIS pointing; this restriction removed SNRs with large spatial extent, such as SN 1006 (e.g., Long et al. 2003). Additionally, we exclude SNRs that are interacting with or are distorted by molecular clouds (e.g., N63A: Chu \& Kennicutt 1988, Warren et al. 2003; G349.7$+$0.2: Lazendic et al. 2005) and those whose large-scale morphologies are the result of axisymmetric winds from pulsars (such as 3C 58, Slane et al. 2004). 

Each source was observed for $\sim$30--1000 ks. Data reduction and analysis was performed using the {\it Chandra} Interactive Analysis of Observations ({\sc ciao}) Version 4.0. We followed the {\sc ciao} data preparation thread to reprocess the Level 2 X-ray data, and we extracted global X-ray spectra of each source using the {\sc ciao} command {\it specextract}. Then, we produced exposure-corrected images of the soft X-rays (0.5--2.1 keV; we set the upper limit to 2.1 keV in order to include all of the flux in the Si {\sc xiii} line) and emission lines (listed in Table~2) for each source. For sources with bright pulsars (like G11.2$-$0.3, Kes 73, RCW 103, and G292.0$+$1.8), the pulsar location and extent was identified using the {\sc ciao} command {\it wavdetect} (a source detection algorithm using wavelet analysis; Freeman et al. 2002). We replaced the region identified by {\it wavdetect} with pixel count values selected from the Poisson distribution of the area surrounding the pulsar using the {\sc ciao} command {\it dmfilth}. This process removed the bright pulsars while preserving the morphologies of the diffuse emission surrounding the pulsars. Generally, the removed area of the pulsars was small ($\ls$16 pixels$^{2}$). In sources where the pulsar emission was more extended (e.g., RCW 103 and Kes 73), it was necessary to replace a larger area ($\ls$400 pixels$^{2}$); since these SNRs are shell-like (Green 2009), this procedure did not alter their overall morphology. No other point sources were removed because of potential confusion with small ejecta substructures. 

Given the young-to-middle age of our sources (see Table~1), we expect that all are ejecta-dominated, and the shocked ISM has only a minor contribution to the observed X-ray flux \citep{bad10}. X-ray spectral modeling generally confirms that abundances are above those of the ISM (see references in Table~1), suggesting the emission is indeed dominated by the shocked SN ejecta and not by shocked ISM. 

\smallskip

\section{Methods}

In what follows, we apply two methods to quantify the X-ray morphologies of our SNR targets: a power-ratio method (PRM) to measure symmetry and wavelet-transform analysis (WTA) to probe X-ray substructure. These techniques were introduced in Lopez et al. 2009a (L09a, hereafter); we refer the reader to that paper for a detailed formalism. Here, we give a brief overview of the methods and their uses. 

\subsection{Power-Ratio Method}

The PRM enables the measurement of asymmetries in X-ray surface brightness distributions, and we employ this technique here to compare the global morphologies of thermal emission in Type Ia and CC SNRs. The method was first applied to characterize the X-ray morphology of galaxy clusters observed with {\it  ROSAT} (Buote \& Tsai, 1995, 1996) and with {\it Chandra} \citep{j05}. Subsequently, L09a and Lopez et al. 2009b (hereafter, L09b) developed and extended the technique to {\it Chandra} observations of SNRs. The PRM measures asymmetries in an image via calculation of the multipole moments of the X-ray surface brightness in a circular aperture. It is derived similarly to the multipole expansion of the two-dimensional gravitational potential within an enclosed radius $R$:

\begin{eqnarray}
\lefteqn{\Psi(R,\phi) = -2Ga_0\ln\left({1 \over R}\right)-2G }
\nonumber \\ & & \times \sum^{\infty}_{m=1} {1\over m
  R^m}\left(a_m\cos m\phi + b_m\sin
m\phi\right), \label{eqn.multipole}
\end{eqnarray}

\noindent
where the moments $a_m$ and $b_m$ are
\begin{eqnarray}
a_m(R) & = & \int_{R^{\prime}\le R} \Sigma(\vec x^{\prime})
\left(R^{\prime}\right)^m \cos m\phi^{\prime} d^2x^{\prime}, \nonumber \\
b_m(R) & = & \int_{R^{\prime}\le R} \Sigma(\vec x^{\prime})
\left(R^{\prime}\right)^m \sin m\phi^{\prime} d^2x^{\prime}, \nonumber
\end{eqnarray}

\noindent
$\vec x^{\prime} = (R^{\prime},\phi^{\prime})$, and $\Sigma$ is the surface mass density. For our imaging analyses, the X-ray surface brightness replaces surface mass density in the power ratio calculation.

The powers of the multipole expansion are obtained by integrating the magnitude of $\Psi_m$ (the \textit{m}th term in the multipole expansion of the potential) over a circle of radius $R$,

\begin{equation}
P_m(R)={1 \over 2\pi}\int^{2\pi}_0\Psi_m(R, \phi)\Psi_m(R, \phi)d\phi.
\end{equation}

\noindent
Ignoring the factor of $2G$, this equation reduces to

\begin{eqnarray}
P_0 & = & \left[a_0\ln\left(R\right)\right]^2 \nonumber \\
P_m & = & {1\over 2m^2 R^{2m}}\left( a^2_m + b^2_m\right) 
\end{eqnarray}

The moments $a_{m}$ and $b_{m}$ (and consequently, the powers $P_{m}$) are sensitive to the morphology of the X-ray surface brightness distribution, and higher-order terms measure asymmetries at successively smaller scales relative to the position of the aperture center (the origin). To normalize with respect to flux, we divide the powers by $P_{0}$ to form the power ratios, $P_m/P_0$. $P_{1}$ approaches zero when the origin is placed at the surface-brightness centroid of an image, so we have set the aperture center in all analyses to the full-band (0.5--8.0 keV) centroid of each remnant. In this case, morphological information is given by the higher-order terms. $P_{2}/P_{0}$ is the quadrupole ratio; examples of sources that have high $P_{2}/P_{0}$ are those with elliptical/elongated morphologies or those with off-center centroids because one side is substantially brighter than the other. $P_{3}/P_{0}$ is the octupole ratio; examples of sources that have high $P_{3}/P_{0}$ are those that have asymmetric or non-uniform surface-brightness distributions.

A Monte Carlo approach described in L09b is used to estimate the uncertainty in the power ratios. Specifically, the exposure-corrected images (normalized to have units of counts) are adaptively-binned using the program {\it AdaptiveBin} \citep{s01} such that all zero pixels are removed, smoothing out noise. Then, noise was put back in by taking each pixel intensity as the mean of a Poisson distribution and selecting randomly a new intensity from that distribution. This process was repeated 100 times for each soft-band image, creating 100 mock images per source. The 1-$\sigma$ confidence limits represent the sixteenth highest and lowest power ratio obtained from the 100 mock images of each source.

\begin{figure*}
\epsscale{1.0}
\includegraphics[width=1.0\textwidth]{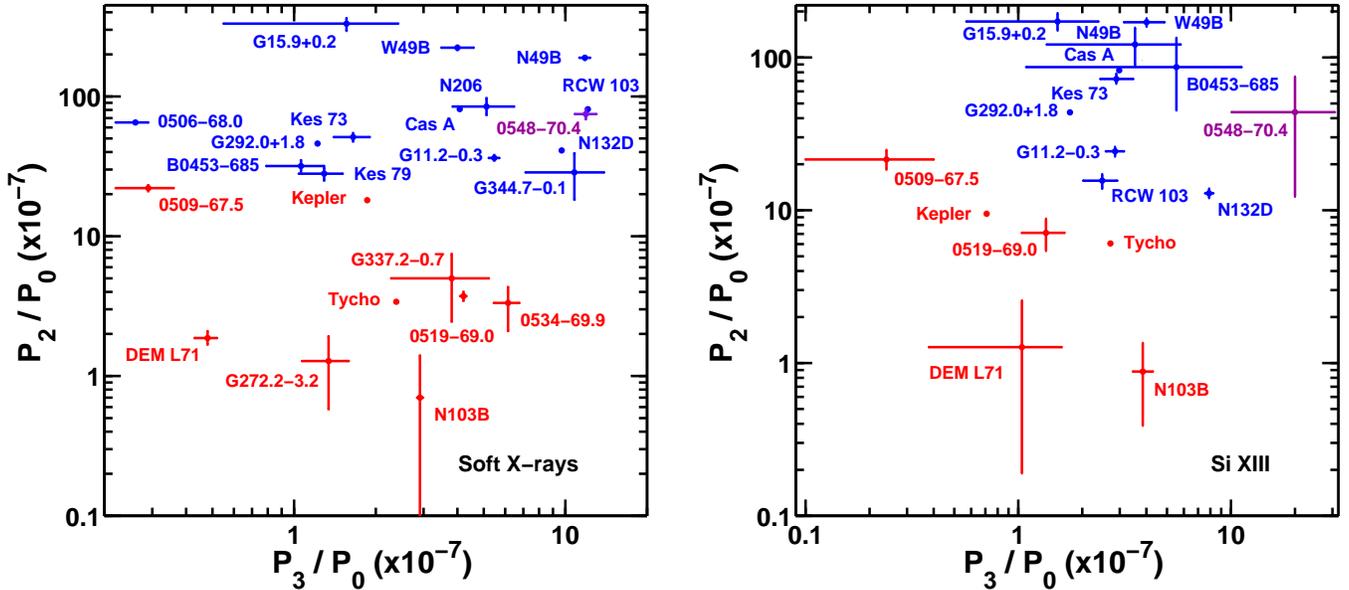}
\caption{{\it Left}: Power ratios, the quadrupole ratio $P_{2}/P_{0}$ versus the octupole ratio $P_{3}/P_{0}$, of the soft X-ray band (0.5--2.1 keV) for twenty-four SNRs in the Milky Way and LMC. {\it Right}: The same plot using only Si {\sc xiii} ($\sim$1.75--2.0 keV) in the seventeen SNRs from L09b.  Type Ia SNRs are in red, CC SNRs are in blue, and 0548$-$70.4 is in purple because of its anomalous ejecta properties that make its type uncertain. The quadrupole ratio is a measure of ellipticity/elongation, and the octupole ratio quantifies the mirror asymmetry of the emission. We find that the Type Ia SNRs are more circular and symmetrical than the CC SNRs.}
\label{fig:p2p3}
\end{figure*}

\subsection{Wavelet-Transform Analysis}

Wavelet-transform analysis (WTA) is the other method we use in this paper to characterize the X-ray morphology of SNRs. In L09a, we demonstrated that this technique can measure accurately the substructure and filling factor of X-ray emitting plasma, and we applied the method to one complex SNR, W49B. Below, we employ WTA to compare the X-ray substructure properties of the line emission in our SNR sample. 

WTA was first applied successfuly to {\it ROSAT} and {\it Einstein} data to extract the small-scale X-ray structure of galaxy clusters \citep{g95}. A wavelet-transformed image is a decomposed image of a signal's intensity (from herein, power) measured at the scale of a filter size. Mathematically, a wavelet transform $w$ is the correlation of a signal $s(x,y)$ in an image with the analyzing wavelet function $g(x,y)$:

\begin{equation}
w(x,y,a) = s (x,y) \otimes \frac{1}{a} g \bigg(\frac{x}{a},\frac{y}{a} \bigg),
\end{equation}

\noindent
where $a$ is the scale (or width) of the wavelet transform. We utilize a radial Mexican-hat function $g(\frac{x}{a},\frac{y}{a})$ (the normalized second derivative of a Gaussian function) of the form

\begin{equation}
g(\frac{x}{a},\frac{y}{a}) = \bigg(2 - \frac{x^{2}+y^{2}}{a^{2}} \bigg) e^{-(x^{2}+y^{2})/2a^{2}}.
\end{equation}

Wavelet-transformed images are produced by calculating $w(x,y,a)$ for each pixel $(m,n)$ in a raw image:

\begin{equation}
w(m,n,a) = \frac{1}{a} \sum c_{ij} g \bigg( \frac{x_i - x_m}{a} , \frac{y_j - y_n}{a} \bigg),
\end{equation}

\noindent
where $c_{ij}$ is the number of counts in the ($i,j$) pixel.

Essentially, $w$ measures the summed intensity enclosed by the area of the Mexican hat. Thus, the size of an individual source can be characterized by the scale where the convolution of the wavelet and a signal reaches a maximum. The wavelet transformation of an isotropic Gaussian signal of size $\sigma$ and intensity $I$ at a position $(x_o,y_o)$ is 

\begin{equation}
w(x_o, y_o, a) = \frac{2 I}{a} \bigg( 1 + \frac{\sigma^2}{a^2} \bigg)^{-2}. 
\end{equation}

\noindent
If we divide this relation by $a$, it has an absolute maximum at $a = \sigma$, which we define as $a_{\rm max}$. Thus, by identifying the peak in a plot of $w/a$ versus $a$ for the central pixel of an emitting substructure, we can measure its size. In addition to measuring the scale of individual substructures, we can sum all the pixels in the wavelet-transformed images at each scale $a$ to find the power profile of an entire source. The resulting power profile, $\langle w \rangle /a$ versus $a$, depends on the scale of isolated structures as well as the filling factor of the emitting material. 

To aid in understanding the method, we describe here how the power profiles would look for a variety of cases. These examples are given quantitatively in $\S$2.3.2 of L09a. In an image with noise only, the power profile would peak at scales of a single pixel and decline rapidly toward zero. If an image has only one substructure without noise, the power profile would have a global maximum at the scale of that substructure. With noise, the power profile should be identical to that of the no-noise case, as long as the signal-to-noise ratio is greater than 2 and the noise-only pixels (those with $a_{\rm max} = 1$ pixel) are removed. As the number of substructures in an image increases, their emission will agglomerate, augmenting the surface area of the emitting regions (i.e., increasing the filling factor) and causing $a_{\rm max}$ of the power profile to increase. 

\section{Results}

We use the methods from $\S$3 to the sample in Table 1 to examine the global and local X-ray morphological properties of SNRs.

\subsection{Global X-ray Morphologies} \label{pr}

To measure the global X-ray morphologies of the thermal emission in SNRs, we applied the PRM to the soft X-ray (0.5--2.1 keV) images of the twenty-four SNRs shown in Figure~\ref{fig:stamps}. This work is an extension of the analyses in L09b, where we employed the PRM and calculated the multipole moments of the Si {\sc xiii} images of seventeen Galactic and LMC SNRs observed by {\it Chandra}. In L09b, we found that the CC and Type Ia SNRs can be distinguished by their quadrupole and octupole ratios, $P_{2}/P_{0}$ and $P_{3}/P_{0}$ respectively. In particular, the CC SNRs had an order of magnitude greater $P_{2}/P_{0}$ than the Type Ia SNRs, indicating CC SNRs are statistically more elongated/elliptical than Type Ia SNRs. Additionally, the CC SNRs had a factor of two larger $P_{3}/P_{0}$ than the Type Ia SNRs, suggesting the CC SNRs are more mirror asymmetric than Type Ia SNRs. The results were the same for other X-ray emission lines besides Si {\sc xiii}, e.g. Ne {\sc ix}, Mg {\sc xi}, and S {\sc xv}. 

\begin{figure*}
\epsscale{1.0}
\includegraphics[width=0.95\textwidth]{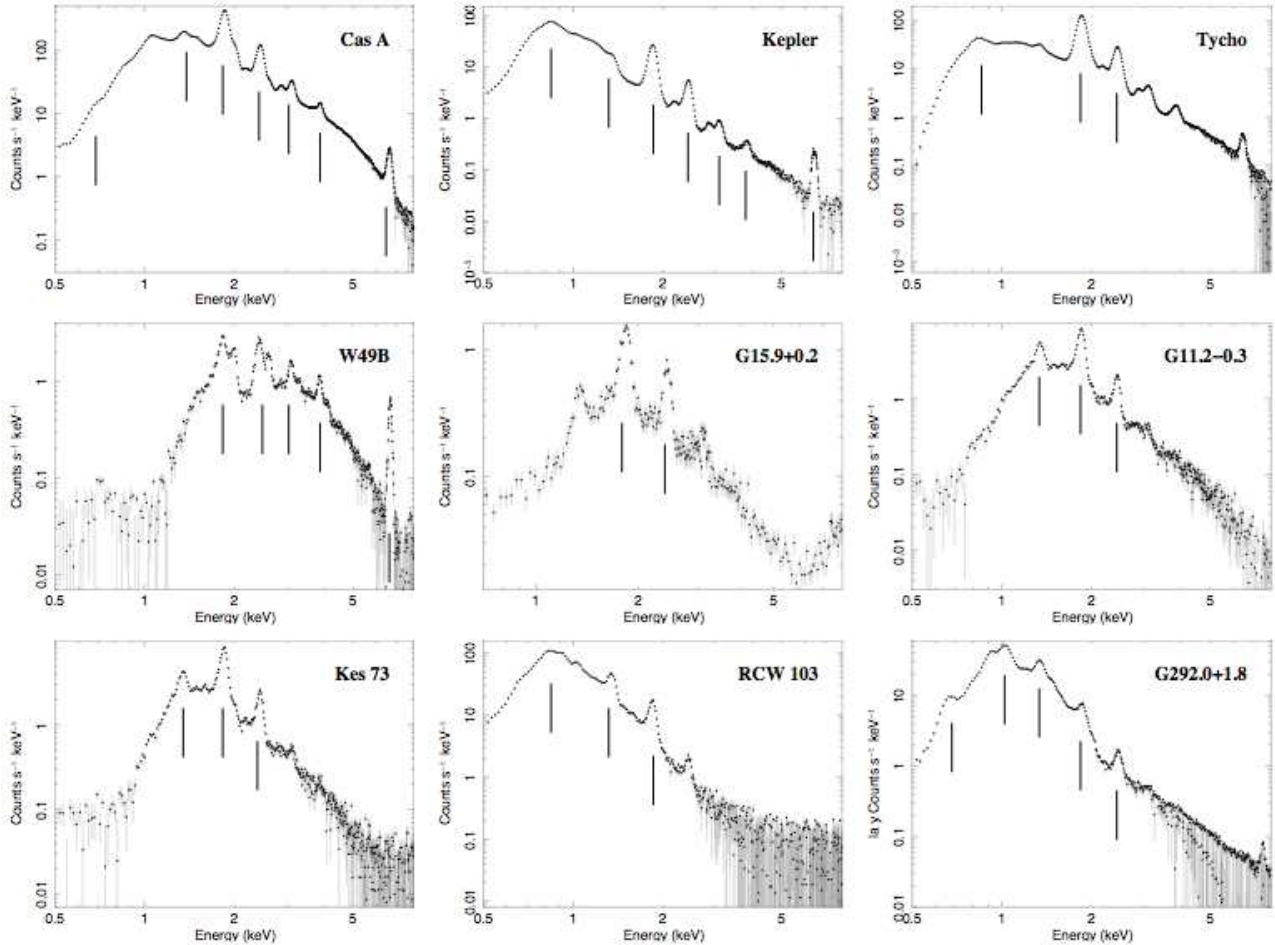}
\caption{{\it Chandra} ACIS full-band (0.5--8.0 keV) spectra for the nine sources analyzed in $\S$4.2. Black lines indicate which emission features we analyzed in $\S$~\ref{wta}; these emission lines are listed in Table 2, from low to high energy.}
\label{fig:spectra}
\end{figure*}

Here, we apply the method to thermal X-rays in SNRs generally. In doing so, we are able to increase the sample size since several remnants have strong bremsstrahlung emission without resolved or strong emission lines. Thus, in addition the seventeen targets from L09b, seven new sources have sufficient bremsstrahlung emission for our analyses: G337.2$-$0.7, G272.2$-$3.2, 0534$-$69.9, 0506$-$68.0, Kes 79, N206, and G344.7$-$0.1. To ensure that we are measuring thermal X-rays and not non-thermal emission, we analyzed the soft X-ray images described in $\S$2, since bremsstrahlung dominates over synchrotron emission below $\sim$2 keV. 

Figure 1 (left) shows the resulting $P_{2}/P_{0}$ versus $P_{3}/P_{0}$ plot for the soft X-ray images; the analogous Si {\sc xiii} plot from L09b is given (right) for comparison. We find that the CC SNRs have a mean $P_{2}/P_{0}$ = (94.2$\pm$0.4)$\times 10^{-7}$ with a standard deviation of 90.4$\times 10^{-7}$, and the Type Ia SNRs have a mean $P_{2}/P_{0}$ = (6.53$\pm$0.05)$\times 10^{-7}$ with a standard deviation of 7.89$\times 10^{-7}$ (excluding SNR 0548$-$70.4, see the discussion in L09b). The mean $P_{3}/P_{0}$ of the two classes are also different: the mean of the Type Ia SNRs is (2.60$\pm$0.13)$\times 10^{-7}$ with a standard deviation of 1.90 and of the CC SNRs is (5.01$\pm$0.88)$\times 10^{-7}$ with a standard deviation of 4.33. This discrepancy in $P_{3}/P_{0}$ can be attributed to the CC SNRs with large $P_{3}/P_{0}$ ($\gs$60). Generally, our findings are consistent with those of L09b: CC SNRs are much more asymmetric or elliptical than Type Ia SNRs. We attribute these differences to the distinct explosion mechanisms and circumstellar medium structures of Type Ia and CC SNRs. 

Of the Type Ia SNRs, Kepler has one of the largest $P_{2}/P_{0}$ because of its off-center centroid (see Figure~\ref{fig:stamps}) since one side being brighter than the other. Sources with more symmetric and homogeneous emission (like G272.2$-$3.2) have the smallest $P_{2}/P_{0}$. Additionally, centrally-filled SNRs (e.g., N103B) tend to have smaller $P_{2}/P_{0}$ as well. Of the CC SNRs, the sources with bright pulsars tend to have the lowest $P_{2}/P_{0}$ (such as B0453$-$685 and Kes 79) suggesting those SNRs are more circular and symmetric than those without pulsars or neutron stars. Finally, SNRs with elongated or elliptical shapes (like W49B) have the highest $P_{2}/P_{0}$, and those with large-scale asymmetries have the greatest $P_{3}/P_{0}$ (e.g., RCW 103). 

\begin{deluxetable}{lc} 
\tablecolumns{2} 
\tabletypesize{\scriptsize}
\setlength{\tabcolsep}{0.0in} 
\tablewidth{0pt}
\tablecaption{X-ray Emission Line Selection}
\tablehead{\colhead{Source} & \colhead{Lines\tablenotemark{a}}}
\startdata
Cas A & O Cont\tablenotemark{b}, Mg {\sc xi}, Si {\sc xiii}, S {\sc xv}, Ar {\sc xvii}, Ca {\sc xix}, Fe {\sc xxv} \\
Kepler & Fe L, Mg {\sc xi}, Si {\sc xiii}, S {\sc xv}, Ar {\sc xvii}, Ca {\sc xix}, Fe {\sc xxv} \\
Tycho & Fe L, Si {\sc xiii}, S {\sc xv} \\
W49B & Si {\sc xiii}, S {\sc xv}, Ar {\sc xvii}, Ca {\sc xix}, Fe {\sc xxv} \\
G15.9$+$0.9 &   Si {\sc xiii}, S {\sc xv} \\
G11.2$-$0.3 & Mg {\sc xi}, Si {\sc xiii}, S {\sc xv} \\
Kes 73 & Mg {\sc xi}, Si {\sc xiii}, S {\sc xv} \\
RCW 103 & Fe L, Mg {\sc xi}, Si {\sc xiii}, \\
G292.0$+$1.8 & O {\sc viii}, Ne {\sc ix}, Mg {\sc xi}, Si {\sc xiii}, S {\sc xv} \\
\enddata
\tablenotetext{a}{Energy ranges for individual lines vary slightly across the sources, depending on e.g., the width of the lines. On average, the bands are: O Cont: 0.6--0.8 keV; O {\sc viii}: 0.6--0.7 keV; Ne {\sc ix}: 0.85--0.95 keV; Fe L: 0.9--1.1 keV; Mg {\sc xi}: 1.20--1.50 keV; Si {\sc xiii}: 1.7--2.1 keV; S {\sc xv}: 2.25--2.60 keV; Ar {\sc xvii}: 2.9--3.3 keV; Ca {\sc xix}: 3.7--4.1 keV; Fe {\sc xxv}: 6.2--6.9 keV.}
\tablenotetext{b}{The oxygen in Cas A is expected to be completely ionized and to dominate the bremsstrahlung continuum \citep{vink96}. Therefore, we use the 0.6--0.8 keV continuum as a proxy for the oxygen.}  
\end{deluxetable}

\subsection{Small-Scale Structure} \label{wta}

From $\S$~\ref{pr}, it is evident that the large-scale morphological differences of the X-ray line and the thermal emitting material {\it between} SNRs can be used to distinguish the explosion type. Next, we consider the relative morphologies of different X-ray lines {\it within} individual sources and what their properties can reveal about their explosions and dynamical evolution. The results of $\S$4.1 hint that the X-ray lines of each SNR have similar morphologies, since the PRM could predict accurately the explosion type regardless of which line image was analyzed.

Toward this end, we apply the WTA technique outlined in $\S$3.2 to all the emission line images of our sources. Since we are focusing on the comparison between emission lines, we limited our sample to only those with at least two strong X-ray line features (with counts per unit area $>$0.01 counts/pixel$^{2}$). Additionally, as we are considering local structures with arcsecond extents, we restrict the analyses of this section to Milky Way SNRs to ensure that we can resolve sub-pc scale structures. Thus, we limit our sample to the nine SNRs that satisfy these criteria. Figure~\ref{fig:spectra} gives the global X-ray spectra for these nine SNRs, with the black lines labeling all the X-ray emission lines whose images we analyzed. Table 2 lists these X-ray emission lines, in order from the lowest to highest energies; the nine SNRs had 2--7 lines that had sufficient surface brightnesses for our analyses.

Figures ~\ref{fig:montagecasa}--\ref{fig:montageg292} (given in the Appendix $\S$A.1) show the raw images of these X-ray lines as well as the resulting wavelet-transformed images at five scales for each SNR (except W49B, given in Figure 14 of L09a). Since the wavelet essentially acts like a filter to pick up the emission at different scales, each field displays the X-ray power of the different lines at the given sizes. At small scales, noise and random fluctuations dominate, and with increasing filter sizes $a$, the distribution and substructure of each ion becomes more evident. Generally, the eight SNRs have similar morphologies among all of their emission lines. These results contrast the case of W49B: in L09a, we found that the iron in W49B was largely absent in the Western half of that source, whereas the other elements were more symmetrically distributed.

\begin{figure*}
\includegraphics[width=0.95\textwidth]{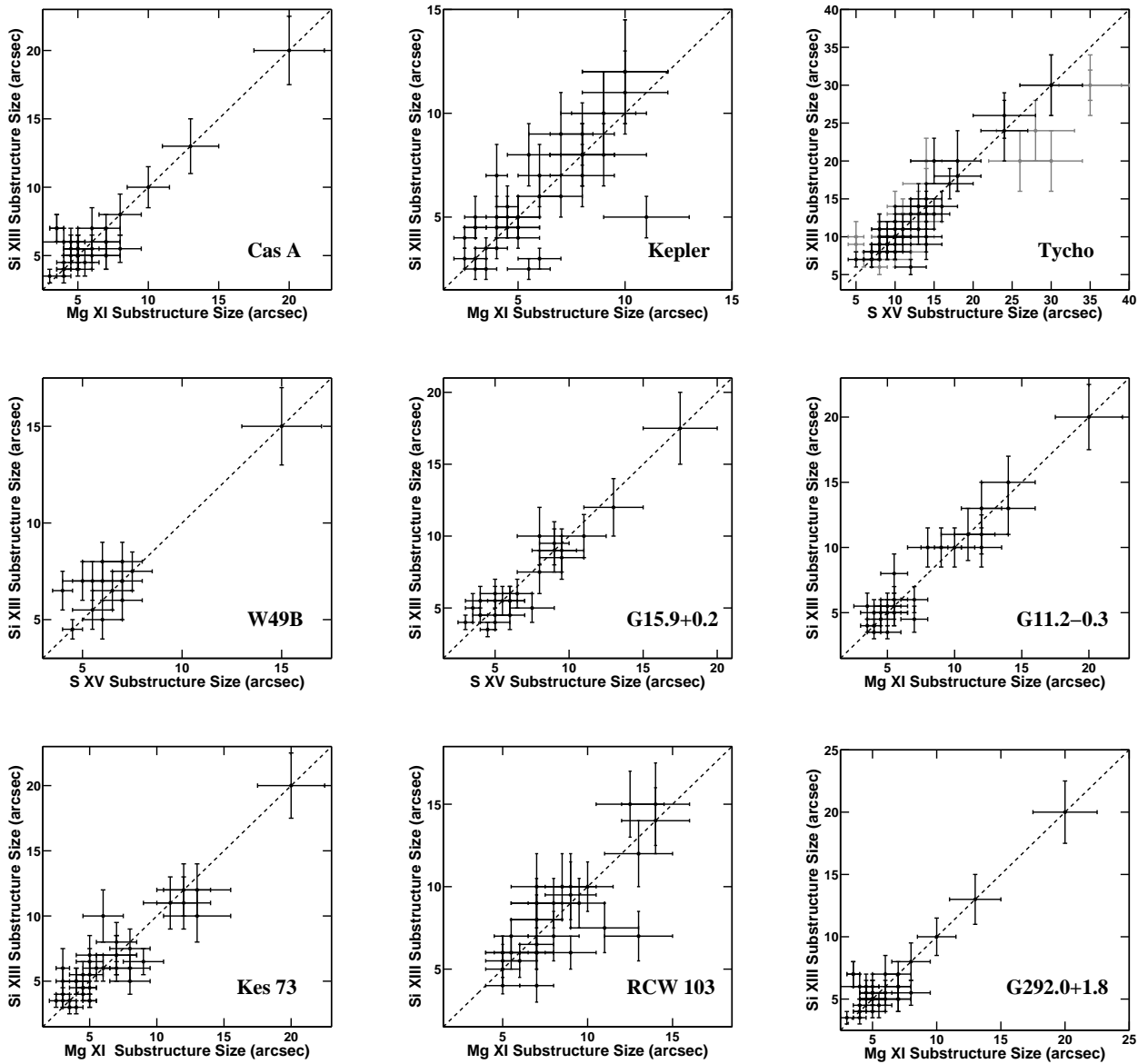}
\caption{Comparison of individual substructure sizes for the two strongest emission lines (Si {\sc xiii} and either Mg {\sc xi} or S {\sc xv}) in each remnant. When available, we compared substructure scales of elements from different burning processes; therefore, Tycho also includes the comparison of Fe L versus Si {\sc xiii} in gray. The error bars reflect the uncertainty in the size estimate, and the dashed lines have slope of unity. 90\% of identified substructures in one X-ray line image had a corresponding substructure in the other X-ray line image of that SNR, and the substructure sizes of the X-ray lines within each source are nearly identical. Only substructures with scales $>$6 pixels were included to avoid PSF effects.}
\label{fig:mgvssi}
\end{figure*}

Using the transformed images, we identify and measure the scale $r_{\rm c}$ and position of individual, isolated X-ray substructures of the line emitting material in each SNR. We find 15--45 substructures in each source, and they span a range of scales ($\sim$3\arcsec--35\arcsec). We only include substructures with scales $\gs$3\arcsec\ to ensure that point-spread-function effects do not influence the results. 

As a probe of chemical mixing, we can compare the substructure sizes $r_{\rm c}$ and locations of different elements within each source. Specifically, we identify and measure substructures which are spatially coincident (defined as those less than ten pixels $\approx$ 5\arcsec~apart) in two X-ray line images. When possible, we restrict our analyses to ions that are the products of different burning processes, and thus the relative scale and position of substructures reveals the effectiveness of chemical mixing in the SNRs. For six sources, we utilize the two strongest X-ray emission lines, Si {\sc xiii} (a product of oxygen burning) and Mg {\sc xi} (a product of carbon and neon burning). In Tycho (which lacks a prominent Mg {\sc xi} feature), we compare the substructures of Si {\sc xiii} and S {\sc xv} (both from oxygen burning) as well as Si {\sc xiii} and Fe L (a product of silicon burning). For the other SNRs (W49B and G15.9$+0$0.3), we are limited to Si {\sc xiii} and S {\sc xv}, since these SNRs have only products of oxygen burning (except for Fe {\sc xxv} in W49B, which has a disparate morphology relative to the other ions; L09a).

Figure~\ref{fig:mgvssi} plots the resulting substructure sizes $r_{\rm c}$ in our nine sources. We find that $\gs$90\% of identified (i.e., the brightest) substructures in one X-ray line image have a corresponding substructure in the other image, suggesting the elements are well mixed throughout the SNRs. Broadly, the slopes of the plots in Figure~\ref{fig:mgvssi} are consistent with unity, indicating that the substructures of the X-ray line emitting material within each SNR have similar physical scales, ranging from 1--16\% of the radius of each SNR. 

\begin{figure*}
\includegraphics[width=0.95\textwidth]{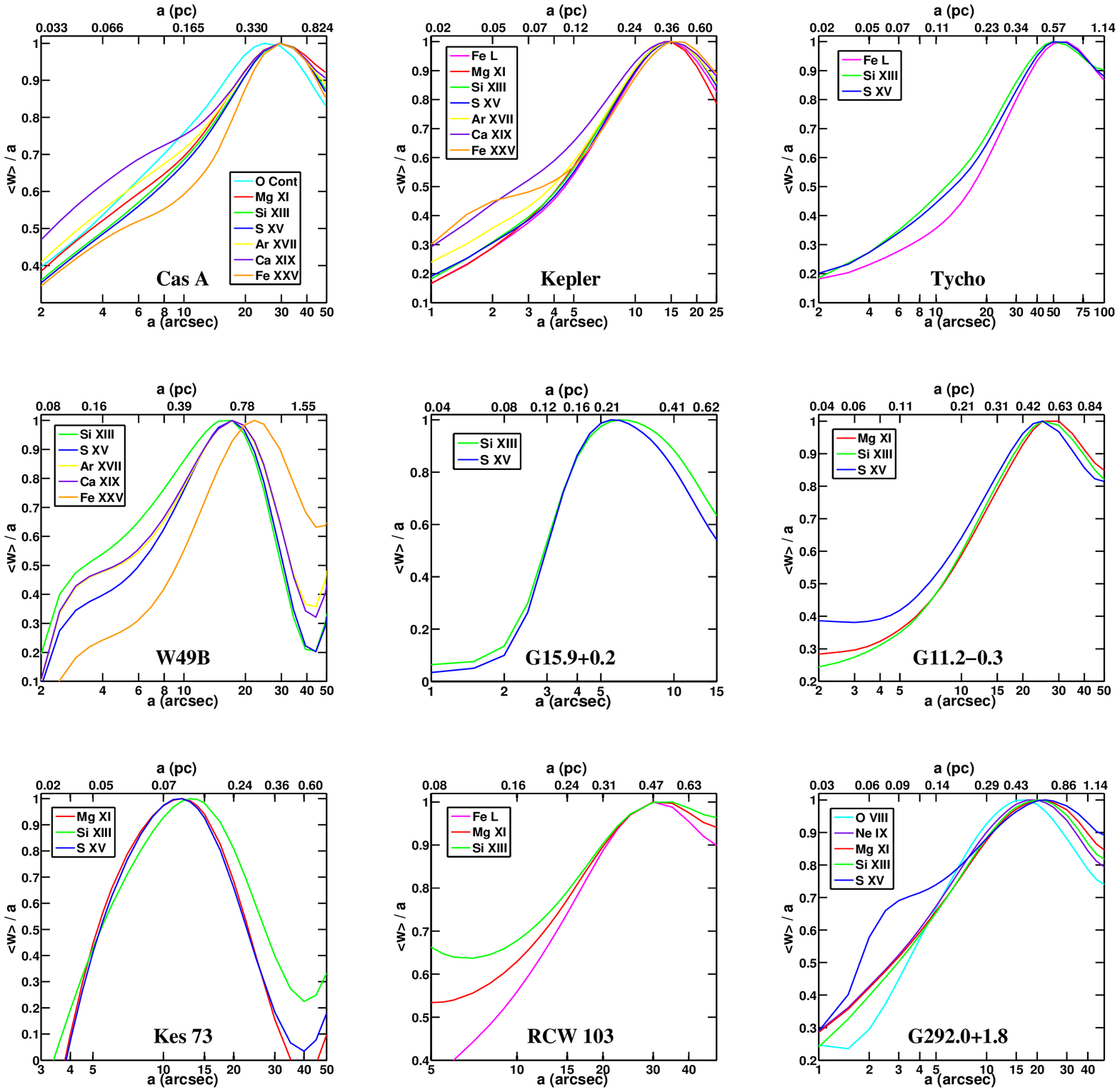}
\caption{Relative power versus substructure scale ($\langle w/a \rangle$ vs. $a$; in arcseconds and parsecs) for all the emission lines in each remnant (see Table 2). The scale $a$ where the power of an individual structure reaches its maximum, $a_{\rm max}$, gives its size. The power profiles of ions in each source have similar shape, with only $\ls$6\% fractional differences in the profiles of a given source. The only exception is W49B, where the Fe {\sc xxv} has substantially less power ($\approx$34\%) at small scales than the other ions.}
\label{fig:walines}
\end{figure*}

\subsection{Average Power over Many Scales}

In addition to comparing individual structures on small scales, we can use WTA to examine the overall power profiles (power as a function of scale over the entire source) of the ions. Figure ~\ref{fig:walines} gives the power profiles ($\langle w \rangle /a$ versus $a$) for the emission lines in each remnant. The curves for each SNR reach maxima at different $a_{{\rm max}}$, ranging from $\sim$10\arcsec--50\arcsec. Since the individual substructures identified above are generally smaller than $a_{\rm max}$ in each source, the scale $a_{\rm max}$ is a reflection of the surface filling factor: the greater the value of $a_{\rm max}$, the larger the filling factor of the emitting material. 

Within each source, the power profiles of the emission lines have similar shape and identical maxima, with only a few exceptions. The Fe {\sc xxv} in W49B peaks at 25\% larger scales than the other ions (22.5\arcsec\ versus 17.5\arcsec) of that source; this discrepancy is the largest among our nine sources. One exception is the O in Cas A, which peaks at $a_{\rm max}=25$\arcsec while the other ions have maxima at $a_{{\rm max}} = 30$\arcsec. Another exception is the Si {\sc xiii} in Kes 73, which peaks at slightly larger scales, $a_{{\rm max}}$ = 13\arcsec, than the Mg {\sc xi} and S {\sc xv}, with $a_{{\rm max}}$ = 12\arcsec. Additionally, some remnants, such as Cas A, Kepler, and G292.0$+$1.8, probably have excess power at small scales in the higher-energy lines (like S {\sc xv}, Ar {\sc xvii}, and Ca {\sc xix}) because of contamination from non-thermal emission. 

To test whether these differences in the ion power profiles are significant, we compare the curves for each SNR quantitatively by measuring their cumulative power across a range of sizes. In particular, we determine the fraction of total power that each ion image has at scales above and below their $a_{\rm max}$ values. For this analysis, in sources where the ions have different $a_{\rm max}$, we used the $a_{\rm max}$ of Si {\sc xiii}. We find that all the SNRs except W49B have $<$6\% differences between their ions' relative power above and below scales of $a_{\rm max}$.  By constrast, the Fe {\sc xxv} in W49B has $\approx$34\% less power than Si {\sc xiii} below scales of its $a_{\rm max} \approx 17.5$\arcsec. We conclude that variations in the SNRs' power profiles are minor (excluding W49B), and the elements of our sources have similar surface filling factors.

\section{Substructure Trends Across SNRs} 

\begin{figure*} 
\includegraphics[width=0.95\textwidth]{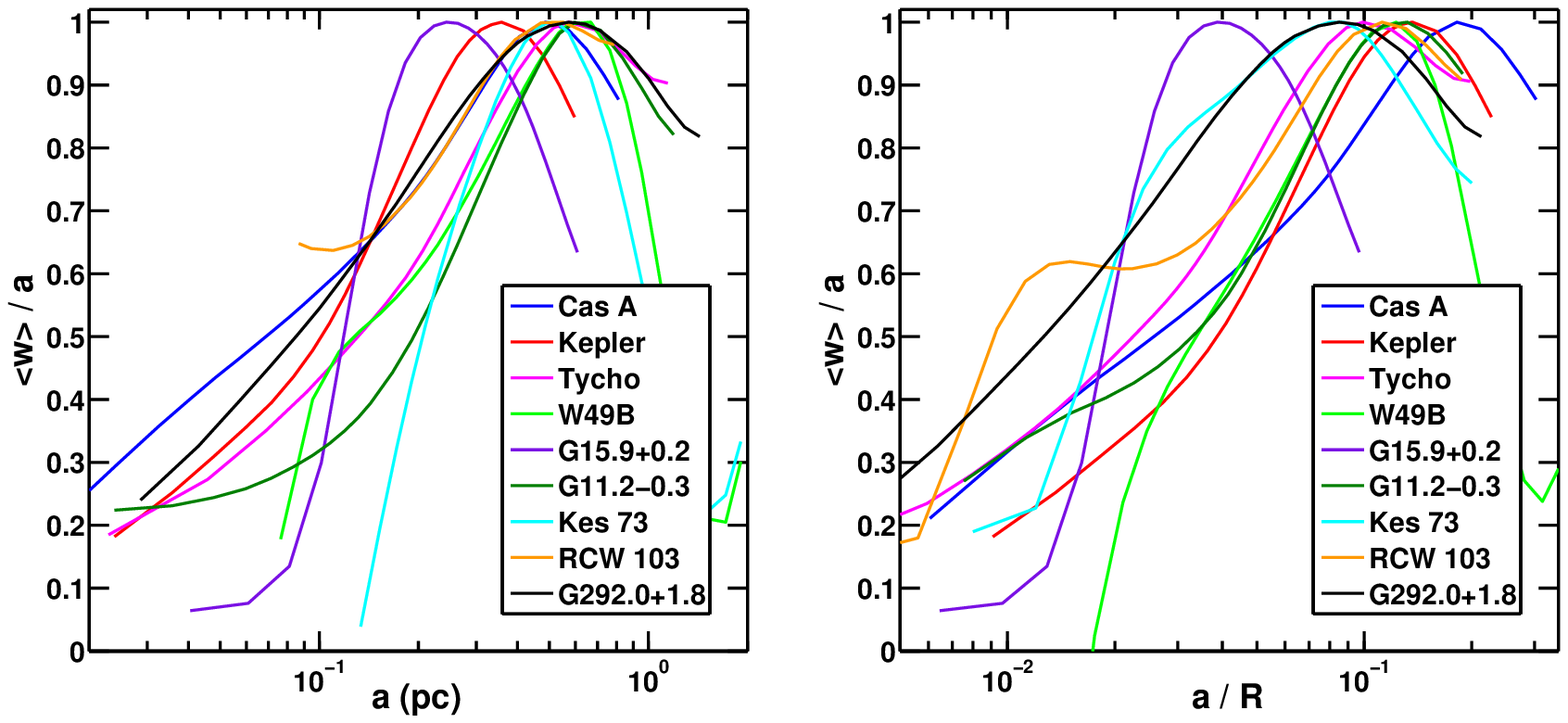}
\caption{{\it Left}: The relative power ($\langle w \rangle /a$) versus scale $a$ in physical units (parsecs). We find that the peak of these curves does not depend on the age of the remnant. {\it Right}: The relative power ($\langle w \rangle /a$) versus the dimensionless quantity $a/R_{\rm X}$, the substructure scale normalized by the size of the remnant. Generally, the peak substructure scale relative to the SNR radius $a_{\rm max}/R_{\rm X}$ is smaller for older sources, indicating that the filling factor of the emitting material decreases with age.}
\label{fig:wa_together}
\end{figure*}

Here, we examine how these substructure characteristics vary across SNRs. First, we compare the power profiles between SNRs to examine whether they depend on age. Toward this end, we plot $\langle w \rangle /a$ of Si {\sc xiii} in every source versus scale $a$ in physical units (parsecs), as shown in Figure ~\ref{fig:wa_together} (left). The $a_{\rm max}$ values vary by only a factor of $\sim$3, from $\approx$0.24 pc for G15.9$+$0.2 to $\approx$0.67 pc for W49B. This value does not appear to depend on age: for example, Tycho has roughly the same maximum as G292.0$+$1.8, $a_{\rm max} \approx$ 0.57 pc, even though G292.0$+$1.8 is almost a factor of ten older than Tycho. The uncertain distances of Kepler and G15.9$+$0.2 may be the reason those two sources are outliers from the other SNRs. 

Since the conversion of scale $a$ to physical units depends on the sometimes largely uncertain distance to our sources, in Figure ~\ref{fig:wa_together} (right), we also plot the power profiles as a function of the dimensionless quantity $a/R_{{\rm X}}$. Here, $R_{\rm X}$ is the radius of the X-ray emission in the full-band image. In elongated sources (such as W49B), we define the radius $R_{\rm X}$ as the semimajor axis of the ellipse that encloses the remnant's full-band X-ray surface brightness. The SNRs' maxima $a_{\rm max}/R_{\rm X}$ span a decent range, from $a_{\rm max}/R_{{\rm X}} \approx$ 0.04 (G15.9$+$0.2) to $a_{\rm max}/R_{{\rm X}} \approx$ 0.57 (G292.0$+$1.8). 

Broadly, the younger SNRs of our sample have larger $a_{{\rm max}}/R_{{\rm X}}$ than the older sources, suggesting that the filling factor of the emission decreases with age. Some remnants (e.g., Tycho and G15.9$+$0.2) do not follow this trend, however. The explanation for this anomaly is uncertain, but we note that Tycho and G15.9$+$0.2 have the shortest ionization timescales $n_{\rm e}t$ (defined as the product of the electron density $n_{\rm e}$ with the time since the plasma was shocked) of our nine sources: Tycho has a mean $n_{\rm e}t \sim 3 \times 10^{10}$ s cm$^{-3}$ (based on our spectral analysis described below) and G15.9$+$0.2 has $n_{\rm e}t \sim 6 \times 10^{10}$ s cm$^{-3}$ \citep{reynolds06}.

Next, we investigate the relationship between individual substructure size and luminosity. To obtain the substructures' emitted fluxes (i.e., absorption corrected), we extracted and modeled the {\it Chandra} X-ray spectra of every substructure identified in the analysis from Figure~\ref{fig:mgvssi}. Spectra were extracted from all available observations with regions of radii corresponding to the scale identified by the WTA. For all the sources except W49B, we fit these spectra using an absorbed, variable abundance plane-parallel shocked plasma model with constant temperature, {\it phabs} $\times$ {\it vpshock}, in XSPEC Version 12.4.0. Previous X-ray analysis of W49B has shown it is in collisional ionization equilibrium (CIE) and requires two plasmas to sufficiently fit its spectra (Miceli et al. 2006; L09a). Therefore, for W49B, we use instead two CIE components for this source: one cool plasma with fixed solar abundances (XSPEC model {\it mekal}) and one hotter plasma with varying supersolar abundances ({\it vmekal}). In the case of Cas A, the continuum emission is thought to be from completely-ionized oxygen rather than H and He \citep{vink96}, so we set the abundances of H and He to zero. Since three sources (G11.2$-$0.3, Kes 73, and RCW 103) do not have published {\it Chandra} X-ray spectra and modeling of their ejecta, we provide a more detailed analysis and discussion of these sources in $\S$A.2. From the best-fit spectral models of the nine SNRs, we determined the emitted flux of each substructure in the Si {\sc xiii} line (over the range 1.75--2.0 keV) and measured the luminosity assuming the distances in Table 1.  

\begin{figure} 
\includegraphics[width=0.95\columnwidth]{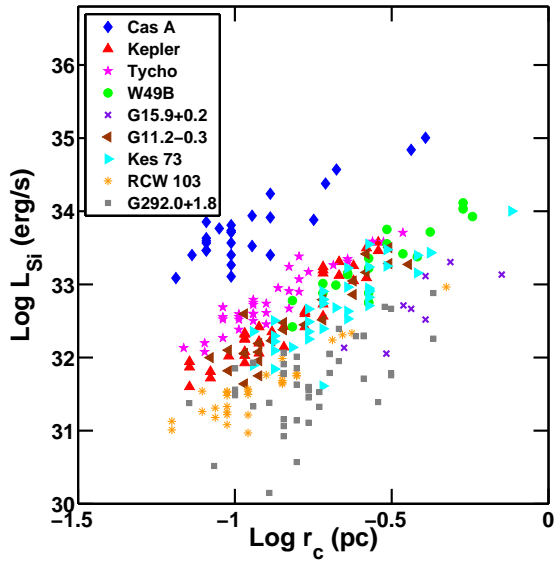}
\caption{Si {\sc xiii} luminosity $L_{\rm Si}$ (1.75--2.0 keV) versus individual substructure size $r_{\rm c}$ for all the emitting regions in the nine SNRs. Best-fit lines to this log-log plot are given in Table 3. We find that substructures in older sources are less luminous than those in younger SNRs. }
\label{fig:Lsi}
\end{figure}

Figure~\ref{fig:Lsi} shows the resulting plot of the Si {\sc xiii} luminosity versus substructure size $r_{\rm c}$ for the nine SNRs. We fit the log-log data with a linear polynomial of the form ${\rm log}~L_{{\rm Si}} = b ({\rm log}~r_{\rm c}) + c$; Table 3 lists the best-fit slopes $b$ and y-intercepts $c$ and associated errors of these analyses. All the slopes $b$ are consistent with values of 2--3. We would expect that if the substructures are optically thin, they would have a power-law index of 3; therefore, we attribute the slopes $1.91<b<2.97$ to a low volume covering fraction of the individual unresolved substructures. In physical terms, the y-intercept $c$ of these fits is the extrapolated luminosity of a substructure that is one parsec in size. Our results show that this parameter decreases with age, spanning over two orders of magnitude from the youngest (Cas A) to oldest (G292.0$+$1.8) of our sources. This trend suggests that individual substructures become less luminous with time, consistent with the result that filling factor decreases with age. 

\begin{deluxetable}{ccc}
\tablecolumns{3} 
\tabletypesize{\footnotesize}
\tablecaption{Best-Fit Lines for Figure~\ref{fig:Lsi}: ${\rm log} L_{{\rm Si}} = b ({\rm log} r_{\rm c}) + c$}
\tablehead{\colhead{Source} & \colhead{$b$} & \colhead{$c$}}
\startdata
Cas A & 2.11$_{-0.48}^{+0.49}$ & 35.8$_{-0.5}^{+0.4}$ \\
Kepler & 2.97$\pm$0.37 & 35.0$_{-0.3}^{+0.4}$  \\
Tycho & 2.39$\pm$0.32 &  34.9$\pm$0.3 \\
W49B & 2.48$_{-0.66}^{+0.65}$ & 34.6$\pm$0.4 \\
G15.9$+$0.9 & 2.53$\pm$1.87 & 33.8$_{-0.9}^{+0.8}$ \\
G11.2$-$0.3 & 2.73$_{-0.52}^{+0.53}$ & 34.7$\pm$0.4 \\
Kes 73 & 2.55$\pm$0.53 & 34.4$_{-0.3}^{+0.4}$ \\
RCW 103 & 2.22$_{-0.36}^{+0.37}$ & 33.6$_{-0.3}^{+0.4}$ \\
G292.0$+$1.8 & 1.91$\pm$0.81 & 33.1$\pm$0.6 \\
\enddata
\end{deluxetable}

We obtain several other physical parameters of the individual substructures with the best-fit spectral models, including their electron temperatures $kT$ and ionization timescales $n_{\rm e}t$. We searched for trends between $kT$ and substructure size as well as $n_{\rm e}t$ and substructure size, and we found no clear relation within nor between SNRs. Since these figures are essentially scatter plots, we do not reproduce them in this paper.

\section{Discussion}

In this paper, we have exploited the wealth of {\it Chandra} ACIS data on galactic and LMC SNRs to examine the observed X-ray properties of these sources. We have applied statistical tools to every remnant with strong line and thermal emission to enable comparison of the local and global morphological characteristics of SNRs. Ultimately, we aimed to determine constraints on the physical processes underlying the dynamical evolution of SNRs. 

In $\S$4.1, we demonstrated that the large-scale morphologies of the X-ray line and thermal emitting material are different for Type Ia versus core-collapse SNRs: the Type Ia SNRs have statistically more spherical and mirror symmetric X-ray emission than the CC SNRs. The ability to distinguish the explosion type based on the bremsstrahlung emission morphology alone enables, for the first time, the typing of remnants with weak X-ray lines and of those with low resolution spectra. It also suggests that it may be possible to type SNRs using other energies where bremsstrahlung dominates. In our analysis, we have successfully identified the SNR G344.7$-$0.1 as originating from a CC explosion, a source of unknown explosion type previously. Additionally, we have confirmed the tentative classifications of G337.2$-$0.7 \citep{rakowski} and of G272.2$-$3.2 \citep{parkaas} as Type Ia SNRs, bringing the number of known Type Ia SNRs in the Milky Way to six (the four others being G1.9$+$0.3: Reynolds et al. 2009; Tycho; Kepler; and SN 1006). 

In $\S$4.2, we investigated the small-scale structures of several ions within nine galactic SNRs. We found that the emission lines within the SNRs have remarkably similar substructure locations, scales, and power profiles, even if the ions are products of different burning processes. This result implies that the metals within the remnants (both in the ejecta and shocked CSM) must have similar spatial distributions, and as such, the metals within SNRs must be globally well-mixed. These findings reinforce observationally that hydrodynamical instabilities efficiently mix ejecta at the scales we can resolve here (although the metals may still be dynamically distinct; see Badenes et al. 2005). Our analysis shows that these results are true for both Type Ia and CC SNRs, indicating that the mixing efficiency is not dependent on explosion type or on CSM structure. Therefore, we conclude that the relative, large-scale morphologies between the different X-ray emission lines in a source cannot be used to distinguish explosion type. 

From our analyses, we do not find evidence of significant ejecta stratification on the scales studied here ($\sim$3\arcsec--40\arcsec) in our Type Ia SNRs, Kepler and Tycho. Chemical stratification is observed in Type Ia SN (e.g., Mazzali et al. 2007), and in some Type Ia SNRs (e.g., Kosenko et al. 2010). Previous X-ray studies of Kepler with {\it XMM-Newton} have shown that Si K and Fe L have similar radial profiles in the north, whereas the Si K extends to larger radii than Fe L in the south \citep{cc}. Indeed, at the smallest scales in Figure~\ref{fig:montagekepler}, the transformed images of Fe L show some substructures interior to those of Mg {\sc xi} and Si {\sc xiii} in the south. In the case of Tycho, prior work has demonstrated that silicon and iron are similarly distributed, with both close to the forward shock \citep{warren05,bad06}. Therefore, our results are consistent with the emerging picture for these two sources: some mechanism (such as hydrodynamical instabilities) has reduced the initial stratification of the ejecta expected right after their explosions.

From our analyses, the only exception to the above statements regarding substructure and mixing is W49B. In that SNR, we found that the Fe {\sc xxv} had distinct morphological and substructure properties from the lower-Z ions (L09a). In L09a, we demonstrated that W49B likely originates from a jet-driven/bipolar explosion (e.g., Ramirez-Ruiz \& MacFadyen 2010), based on these morphological discrepancies and the abundance ratios and masses of the different species. Although Cas A also has jet-like features, these structures do not contribute significantly to the remnant's overall X-ray surface brightness, and the power profiles of all the ions in Cas A are identical. Thus, we conclude that only large-scale discrepancies between ions' spatial distributions within a source can facilitate identification of the explosion mechanism.

Our findings reinforce the unique nature of W49B as the remnant of a jet-driven explosion. We can estimate whether this kind of event is expected to have occurred recently in the Milky Way (MW) galaxy by considering the observed rates of SNe in the Local Group. Jet-driven/bipolar explosions are associated with Type Ib/Ic SNe, a subclass of CC SNe. In the MW, the rate of CC SNe is $\sim$2 per century \citep{tammann}. Of these CC SNe, approximately 20\% are Type Ib/Ic \citep{smartt}; thus, the rate of Type Ib/Ic SNe is $\sim$1/250 years. Some subset of these SNe will be jet-driven/bipolar explosions, but it is still uncertain what fraction of Type Ib/Ic are jet-driven. However, the estimated rate of hypernovae (HNe, which are super-energetic bipolar SNe) per galaxy is $\sim 10^{-5}$/year \citep{izz,pod}. If 1--10\% of bipolar explosions are HNe, the rate of jet-driven SNe would be one every 10000--1000 years. Therefore, it is reasonable to expect several jet-driven/bipolar SNe in the MW, and a few should be observable as SNRs at X-ray wavelengths.

From the comparative analysis of our sources, we have set several observational constraints that are useful tools to test the validity of theoretical models of SNR dynamical evolution. Specifically, we have found that: 

\begin{itemize}
\item Mixing of shocked ejecta and CSM must be efficient on the scales resolved here ($\sim$3--35\arcsec);
\item Individual emitting substructures within a source span a range of sizes, and $\gs$90\% of the metals' brightest substructures in an SNR should be coincident and have equal surface emission scales;
\item The surface area of the X-ray emission (given by our parameter $a_{\rm max}$) does not depend on age (likely because it depends on a combination of age and density). However, relative to the size of the remnant, the surface area of the X-ray emission decreases with age;
\item Individual emitting substructures become less luminous with time;
\item The scale of individual substructures does not tightly correlate with the temperature or ionization timescale of that substructure. 
\end{itemize}  

The list above is a first step toward a broad observational baseline for direct comparison to the predictions of hydrodynamical models. As such, the local and global morphological properties described here should aid in advancing understanding of SNRs, both from an observational and theoretical perspective. 

\nocite{badreview,b95,b96,chu88,f02,kosenko,laz,lal2,long03,m33,mazzali,miceli,err,slane04,warren03,wh06}

\acknowledgements

The authors thank Tesla Jeltema for helpful discussions. This work is supported by NASA {\it Chandra} grant ARO-11009X (LAL and ER-R), DOE SciDAC DE-FC02-01ER41176 (LAL and ER-R), a National Science Foundation Graduate Research Fellowship (LAL), an AAUW American Dissertation Fellowship (LAL), and the Packard Foundation (ER-R).

\bibliographystyle{apj}
\bibliography{lines}

\clearpage

\appendix

\section{A.1. Wavelet-Transformed Images}

\begin{figure}[ht!]
\includegraphics[width=0.9\textwidth]{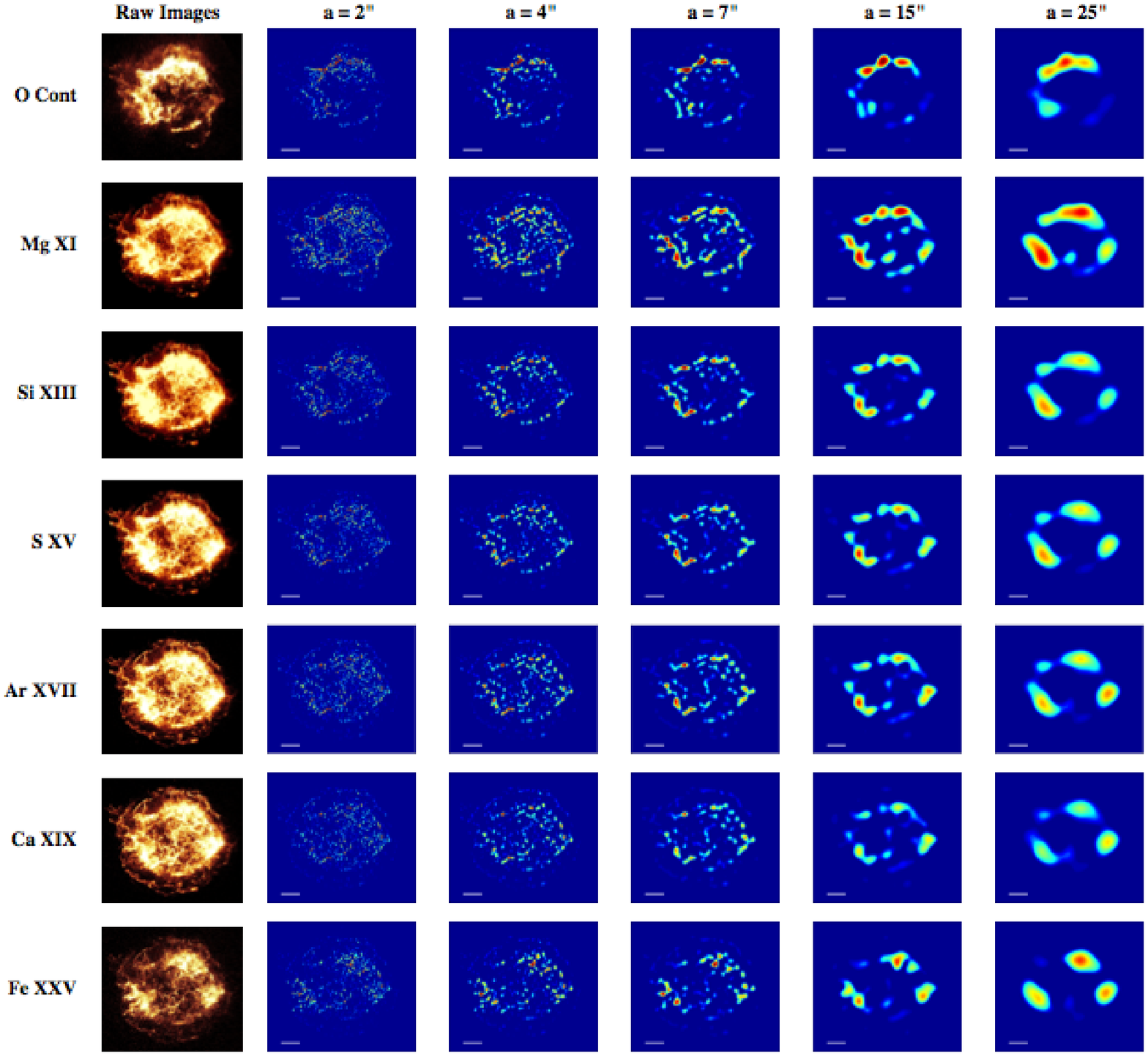}
\caption{Raw images of line emission (O continuum, Mg {\sc xi}, Si {\sc xiii}, S {\sc xv}, Ar {\sc xvii}, Ca {\sc xix}, and Fe {\sc xxv}) in Cas A and corresponding wavelet-transformed images for five different scales. The white scale bar is 1' $\approx$ 1 pc in length. The color bar is set so blue is the minimum, and red is the maximum.}
\label{fig:montagecasa}
\end{figure}

\clearpage

\begin{figure}
\includegraphics[width=0.9\textwidth]{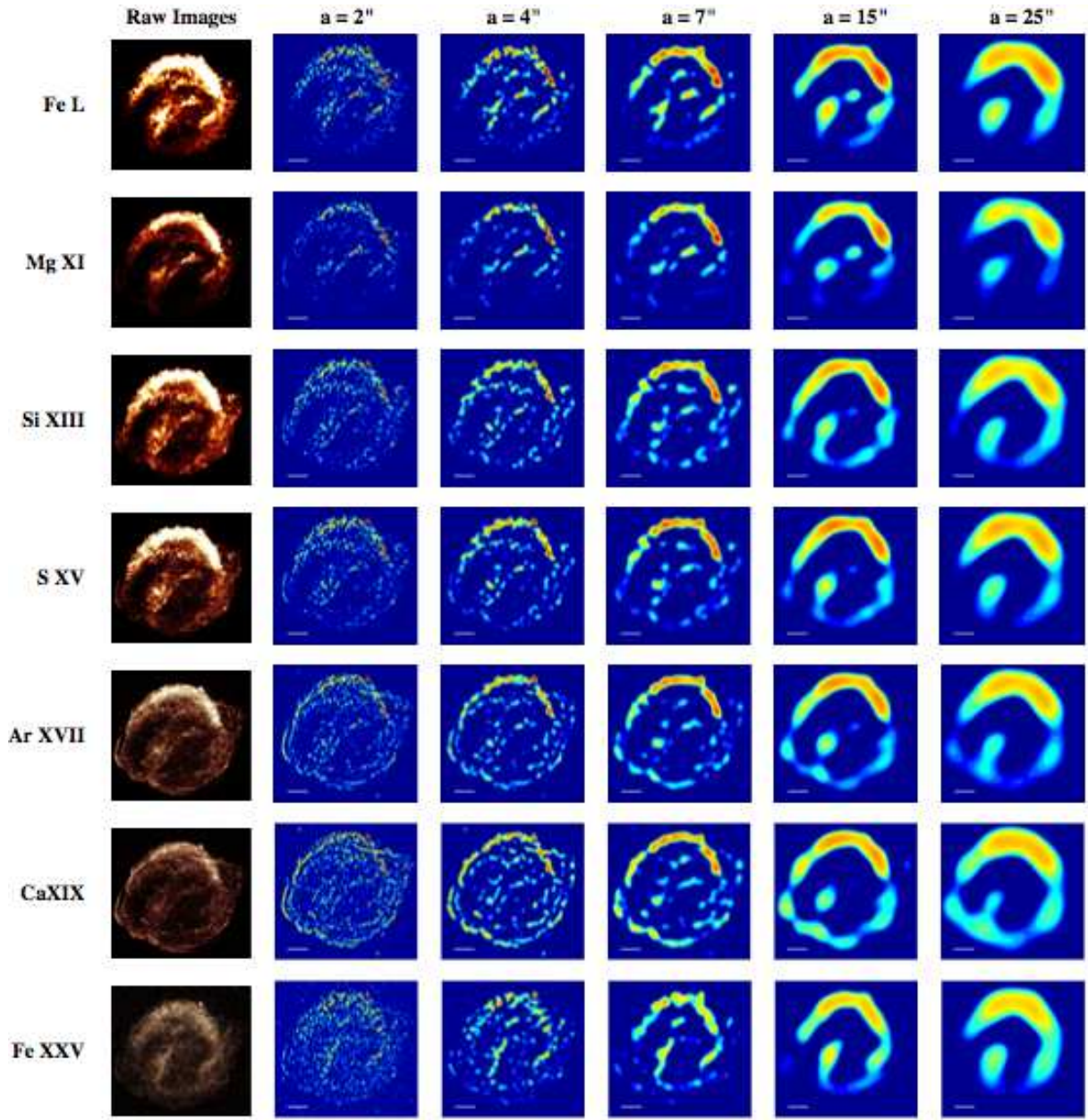}
\caption{Raw images of line emission (Fe L, Mg {\sc xi}, Si {\sc xiii}, S {\sc xv}, Ar {\sc xvii}, Ca {\sc xix} and Fe {\sc xxv}) in Kepler and corresponding wavelet-transformed images for five different scales. The white scale bar is 41\arcsec $\approx$ 1 pc in length.}
\label{fig:montagekepler}
\end{figure}

\clearpage

\begin{figure}
\includegraphics[width=0.95\textwidth]{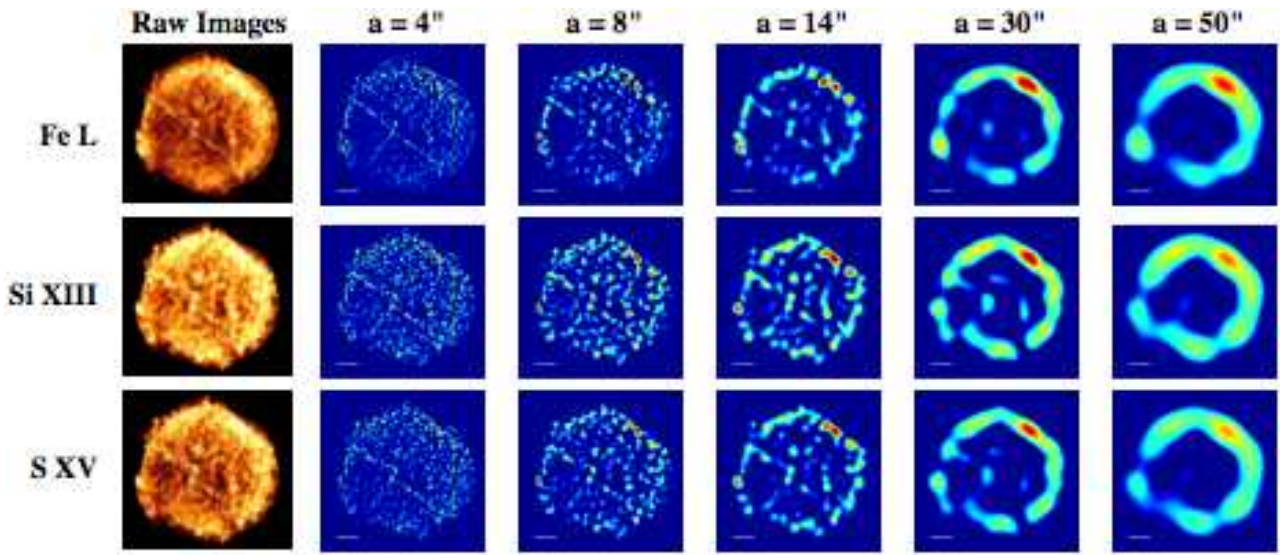}
\caption{Raw images of line emission (Fe L, Si {\sc xiii}, and S {\sc xv}) in Tycho and corresponding wavelet-transformed images for five different scales. The white scale bar is 87\arcsec $\approx$ 1 pc in length.}
\label{fig:montagetycho}
\end{figure}

\begin{figure}
\includegraphics[width=0.95\textwidth]{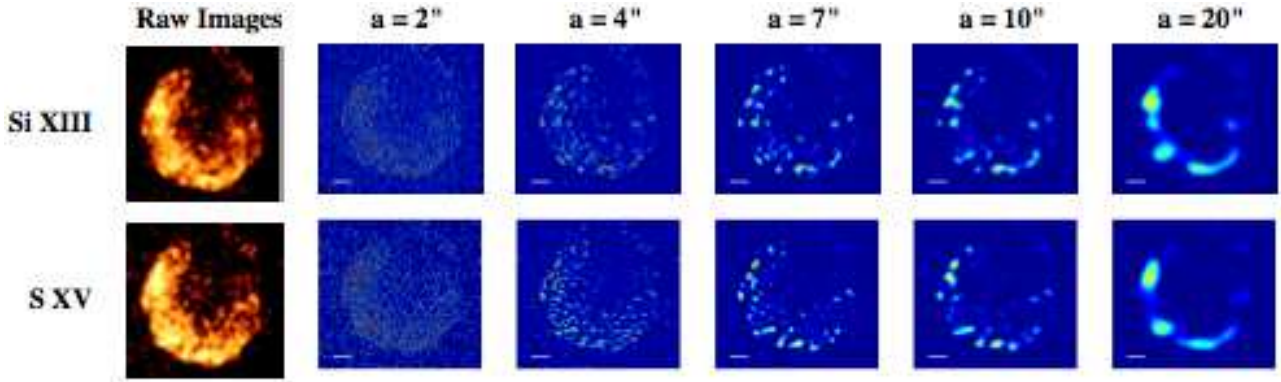}
\caption{Raw images of line emission (Si {\sc xiii} and S {\sc xv}) in G15.9$+$0.3 and corresponding wavelet-transformed images for five different scales. The white scale bar is 50\arcsec $\approx$ 1 pc in length.}
\label{fig:montageg15}
\end{figure}

\begin{figure}
\includegraphics[width=0.95\textwidth]{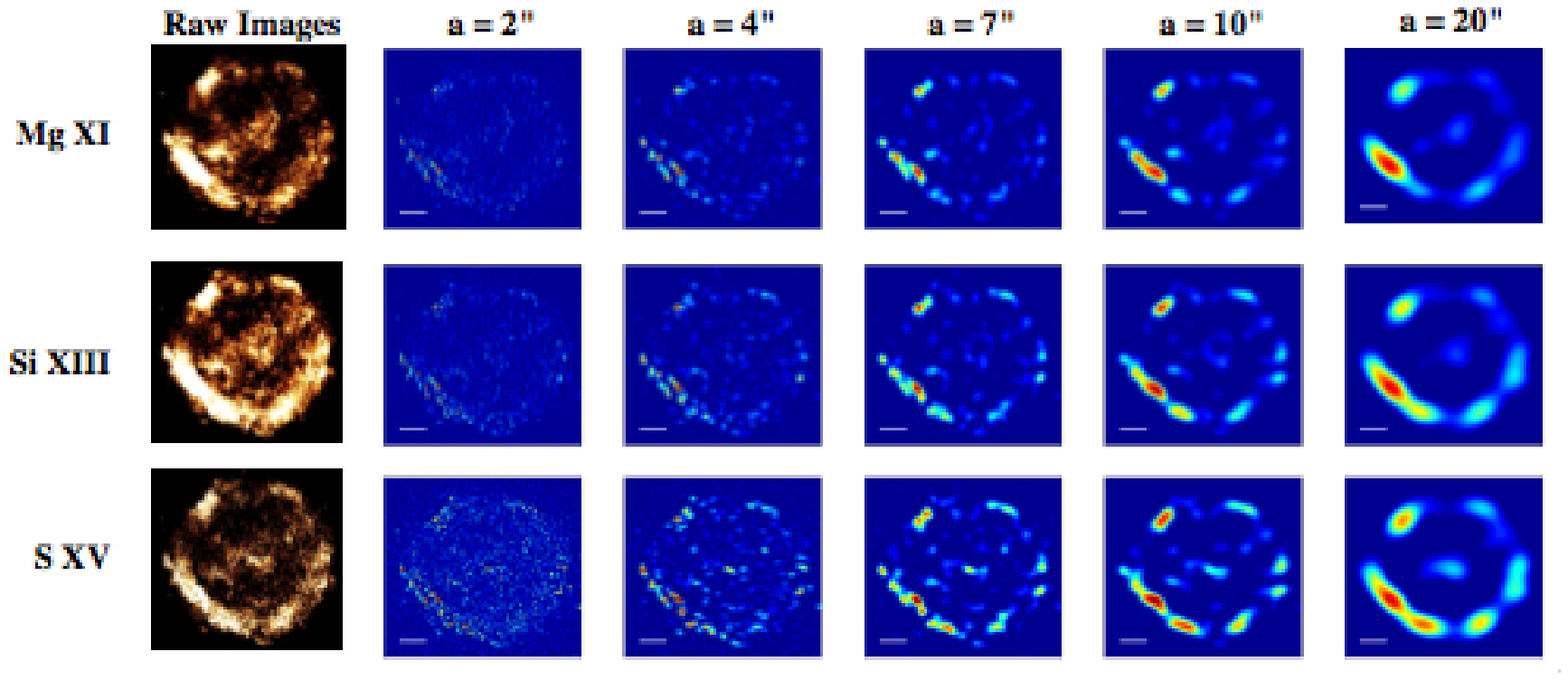}
\caption{Raw images of line emission (Mg {\sc xi}, Si {\sc xiii}, S {\sc xv}) in G11.2$-$0.3 and corresponding wavelet-transformed images for five different scales. The white scale bar is 41\arcsec $\approx$ 1 pc in length.}
\label{fig:montageg11}
\end{figure}

\begin{figure}
\includegraphics[width=0.95\textwidth]{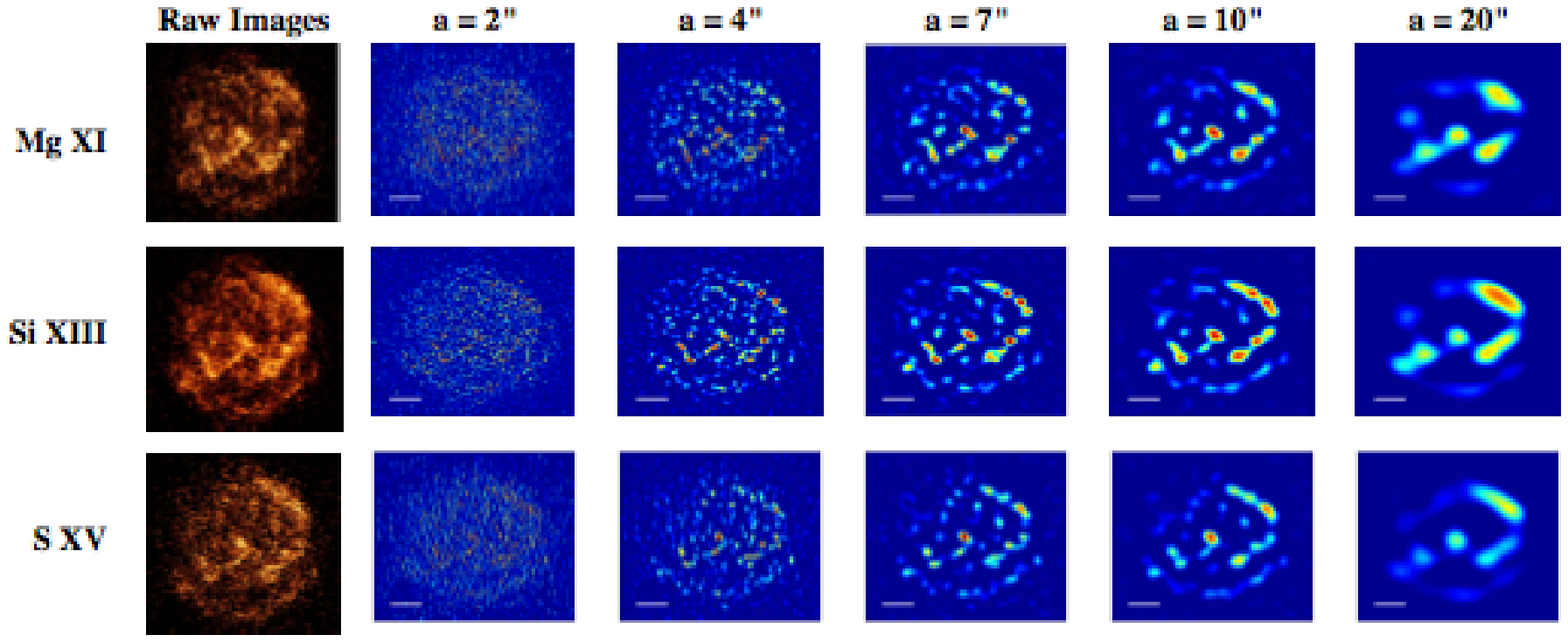}
\caption{Raw images of line emission (Mg {\sc xi}, Si {\sc xiii}, S {\sc xv}) in Kes 73 and corresponding wavelet-transformed images for five different scales. The white scale bar is 52\arcsec $\approx$ 2 pc in length.}
\label{fig:montagekes73}
\end{figure}

\begin{figure}
\includegraphics[width=0.95\textwidth]{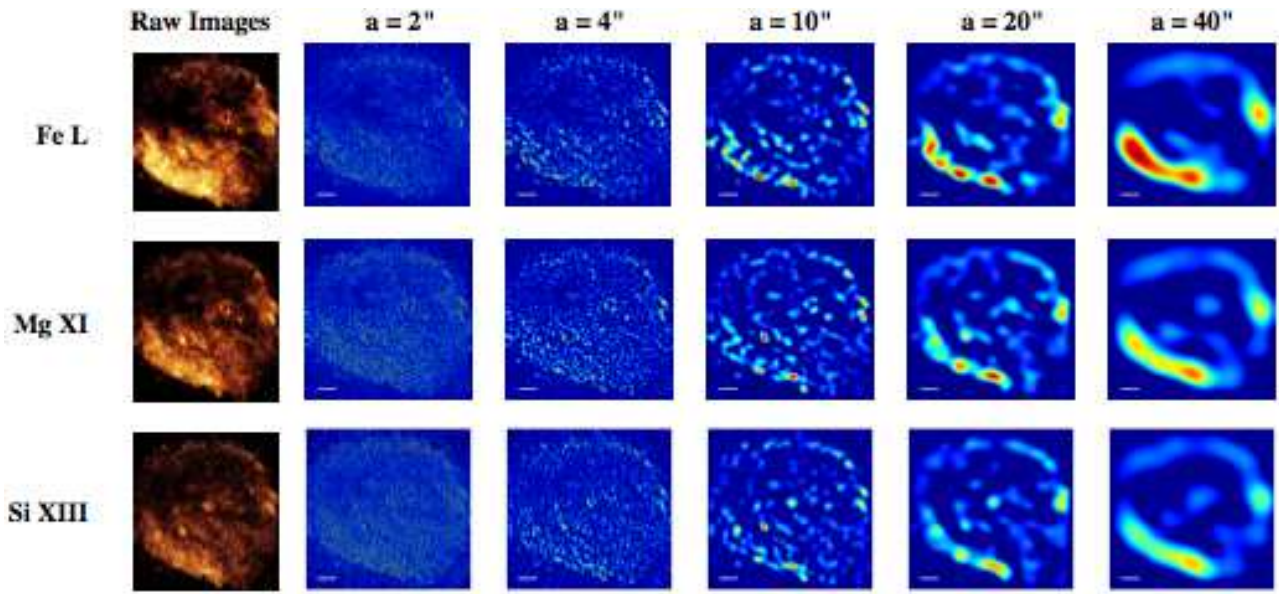}
\caption{Raw images of line emission (Fe L, Mg {\sc xi}, Si {\sc xiii}) in RCW 103 and corresponding wavelet-transformed images for five different scales. The white scale bar is 63\arcsec $\approx$ 1 pc in length.}
\label{fig:montagercw103}
\end{figure}

\clearpage

\begin{figure}
\includegraphics[width=0.95\textwidth]{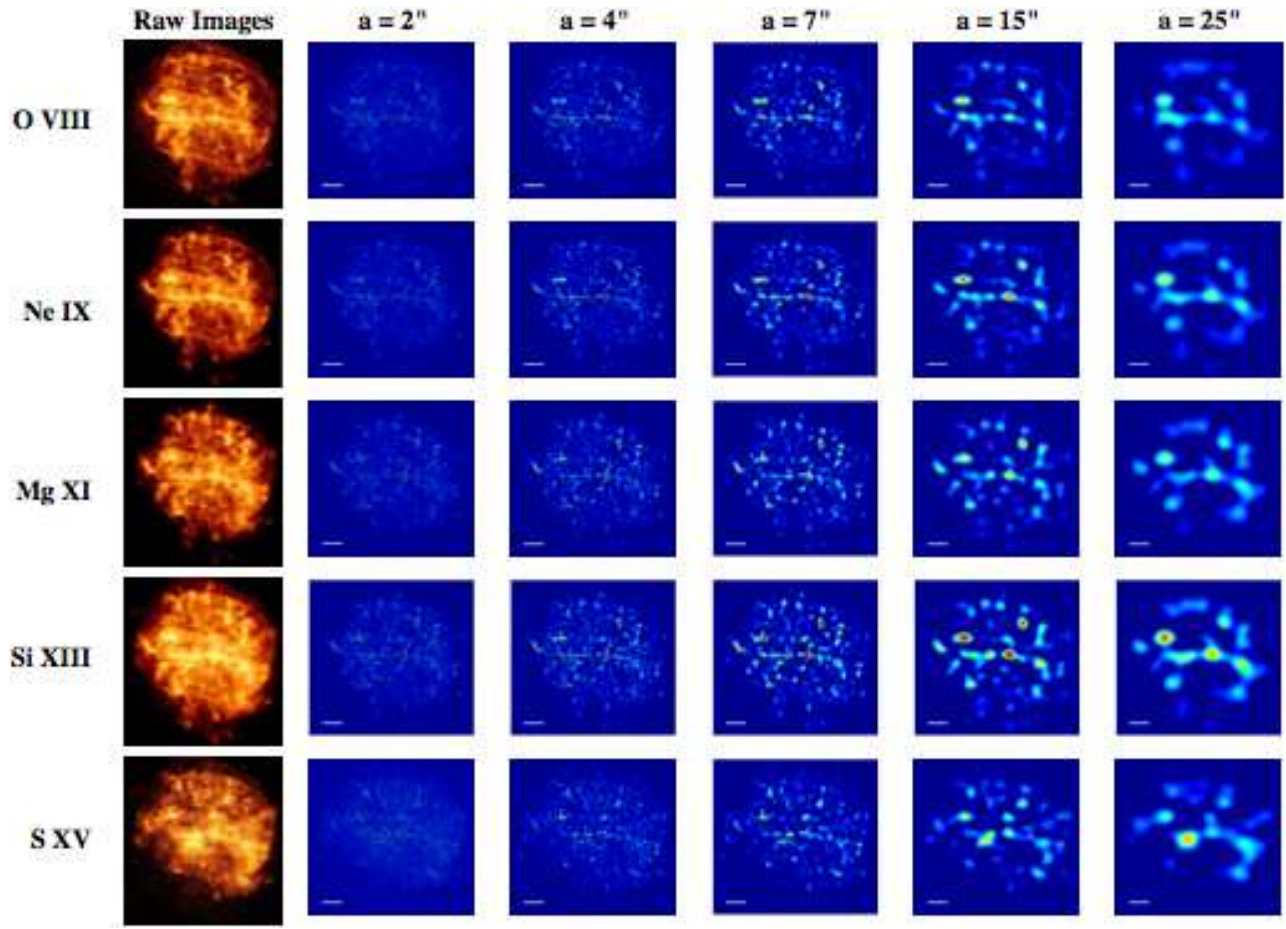}
\caption{Raw images of line emission (Ne {\sc ix}, Mg {\sc xi}, Si {\sc xiii}, S {\sc xv}) in G292.0$+$1.8 and corresponding wavelet-transformed images for five different scales. The white scale bar is 69\arcsec $\approx$ 2 pc in length.}
\label{fig:montageg292}
\end{figure}

\clearpage

\section{A.2. Detailed Spectral Modeling}

Since three of the nine SNRs in our WTA sample (G11.2$-$0.3, Kes 73, and RCW 103) do not have published {\it Chandra} X-ray spectra and modeling of their ejecta, we provide here a more detailed analysis and presentation of the spectra from these targets.  

\begin{figure*} 
\includegraphics[width=\columnwidth]{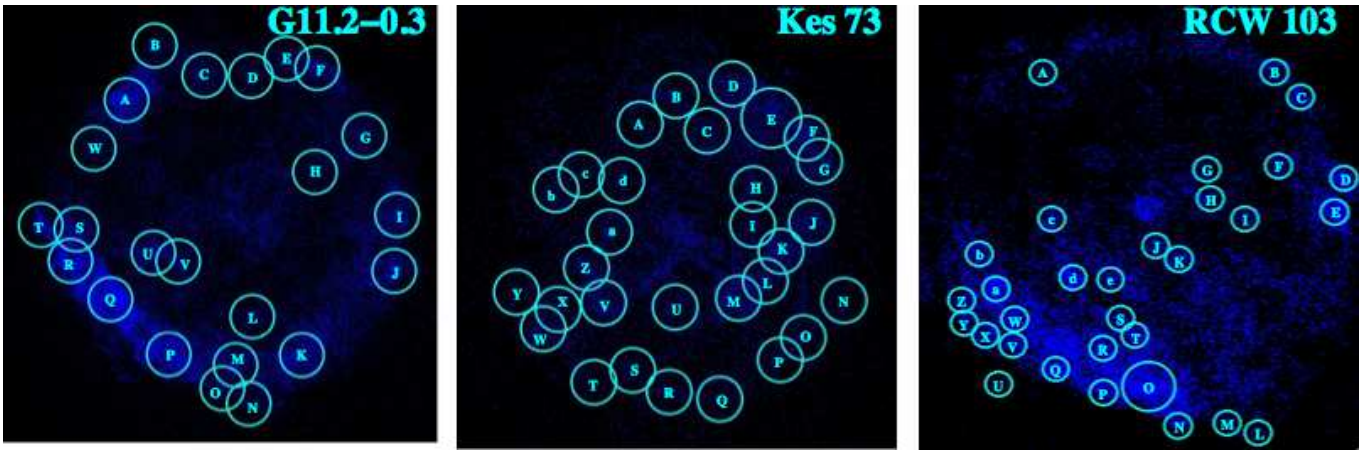}
\caption{Twenty-three substructures in G11.2$-$0.3 (left), thirty substructures in Kes 73 (middle), and thirty-one substructures (right) identified with WTA. We extracted X-ray spectra at these locations, and the fit results are given in Tables 4--6.}
\label{fig:regions_all3}
\end{figure*}

\begin{figure*} 
\includegraphics[width=0.95\textwidth]{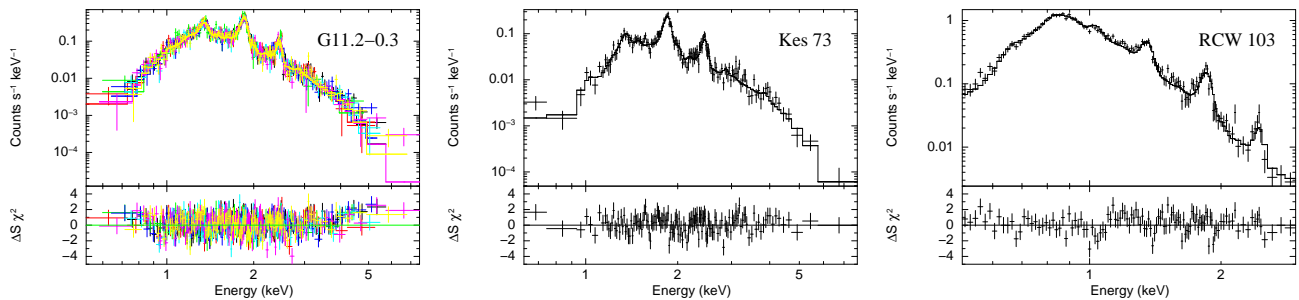}
\caption{Example {\it Chandra} X-ray spectra, models, and residuals from G11.2$-$0.3 (left; from circle Q), Kes 73 (middle; from circle F), and RCW 103 (right; from circle R).}
\label{fig:examples_all3}
\end{figure*}

We identified 23 substructures with WTA in G11.2$-$0.3. We extracted {\it Chandra} spectra for the seven ACIS observations of G11.2$-$0.3 from regions centered on the 23 substructures. Since the goal of this analysis is to describe the physical state of the ejecta, here we chose to improve the statistics by increasing the radii of the regions where we extracted spectra to 30 pixels = 14.76\arcsec (the cyan circles A--W in Figure~\ref{fig:regions_all3}, left). Background spectra were produced from a region $\approx$50\arcsec\ from G11.2$-$0.3 and subtracted from the source spectra. Spectra were modeled like the analyses above, with an absorbed plane-parallel shocked plasma model, and data from the seven observations were fit simultaneously to improve statistics. We let the abundances of magnesium, silicon, sulfur, and iron vary freely with the other elements frozen to solar values in the fits. Example spectra and models from one region (circle Q in Figure ~\ref{fig:regions_all3}, left) are given in Figure ~\ref{fig:examples_all3} (top). 

Table 4 lists the parameters of the best-fit models. The mean absorbing column density is high, $\langle N_{H} \rangle $ = 2.4$\times$10$^{22}$ cm$^{-2}$, and attenuates the soft X-rays below about 1 keV. The electron temperature $kT$ varies from 0.60 keV up to 1.38 keV and has a mean value of $ \langle kT  \rangle $ = 0.80$\pm$0.18  keV. The Northern (regions C and D) and the Western (regions I and J) regions seem to be hotter than other parts of the remnant, and the area just south of the pulsar (regions U, V, and L) appear to have elevated temperatures as well. The abundances of magnesium, silicon, and sulfur are supersolar, consistent with a shocked ejecta origin of the X-ray emission lines. The only exception is region W, which is near solar metallicity, suggesting that the X-ray emission is from shocked ISM. The ionization timescale $n_{e} t$, the product of ambient electron density and the time since the plasma was shock heated, spans an order of magnitude, with $n_{e} t = 9.47 \times 10^{10}- 1.04 \times 10^{12}$ s cm$^{-3}$. The mean ionization timescale from our fits is $ \langle n_{e}t \rangle = 4.85 \times 10^{11}$ s cm$^{-3}$. 

In Kes 73, we identified 30 substructures with WTA. We extracted the {\it Chandra} spectra at these locations with radii of 30 pixels = 14.76\arcsec (cyan circles A--d in Figure~\ref{fig:regions_all3}, middle). Background spectra were produced from a region $\approx$50\arcsec\ from Kes 73 and subtracted from the source spectra. Spectra were modeled as above, with a single plasma in non-equilibrium ionization. We let the abundances of magnesium, silicon, sulfur, and iron vary freely with the other elements frozen at solar values during the fits. An example spectrum and model from one region (circle F in Figure~\ref{fig:regions_all3}, middle) are given in Figure~\ref{fig:examples_all3} (middle).
   
Table 5 lists the parameters of the best-fit models. As in G11.2$-$0.3, the mean absorbing column density is high, $\langle N_{H} \rangle $ = 2.7$\times$10$^{22}$ cm$^{-2}$, and attenuates the soft X-rays below about 1 keV. The $N_{H}$ varies across the remnant, from 2.1--3.4$\times$10$^{22}$ cm$^{-2}$, and appears to be elevated in the northern portions of Kes 73 (e.g., regions C and D). The electron temperature $kT$ ranges from 0.63 keV up to 1.22 keV and has a mean value of $ \langle kT  \rangle $ = 0.84$\pm$0.49 keV. Generally, the abundances of magnesium, silicon, and sulfur are supersolar, with a few exceptions: seven regions (R, S, T, a, b, c, and d) are consistent with solar metallicity within their errors. This result suggests that the southern and Eastern sections of Kes 73 are dominated by X-ray emission from the shocked ISM. The ionization timescale $n_{e} t$ spans nearly two orders of magnitude, with $n_{e} t = 7.75 \times 10^{10}- 5.24 \times 10^{12}$ s cm$^{-3}$. The mean ionization timescale from our fits is $ \langle n_{e}t \rangle = 7.85 \times 10^{11}$ s cm$^{-3}$. 

In RCW 103, we identified 31 substructures with WTA. We extracted the {\it Chandra} spectra at these locations with radii of 30 pixels = 14.76\arcsec (cyan circles A--e in Figure~\ref{fig:regions_all3}, right). Region O was larger, with a radius of 60 pixels since its $a_{{\rm max}}$ was large, $\approx$60 pixels. Background spectra were produced from a rectangular region with $\approx$150$\times$15\arcsec\ sides south of RCW 103 (chosen to avoid chip gaps) and subtracted from the source spectra. Spectra were modeled as above but excluding energies above 3 keV because of a dominant non-thermal component. We let the abundances of magnesium, silicon, and iron vary freely with the other elements frozen at their solar values during the fits. An example spectrum and model from one region (circle F in Figure~\ref{fig:regions_all3}) are given in Figure~\ref{fig:examples_all3} (bottom).
   
Table 6 lists the parameters of the best-fit models. These fits were successful statistically, except for one region, circle O in Figure~\ref{fig:regions_all3}, which was poorly fit by one NEI plasma. The addition of a second NEI plasma improved that fit statistically, with a $\Delta \chi^2 =$ 58, for 126 degrees of freedom. The mean absorbing column density of our RCW 103 spectral models is moderate, $\langle N_{H} \rangle $ = 5.4$\times$10$^{21}$ cm$^{-2}$, so the soft X-rays are relatively unattenuated and the broad Fe L emission is strong. The electron temperature $kT$ is fairly constant across the remnant, with all models giving $kT \approx$ 0.52--0.67 keV except one, region L, at 0.33 keV. The mean electron temperature is $ \langle kT  \rangle $ = 0.56$\pm$0.14 keV. The iron abundances have roughly solar values, suggesting the Fe L emission comes from the shock-heated ISM. The magnesium and silicon have supersolar abundances, with only a couple outliers, so the Mg {\sc xi} and Si {\sc xiii} lines likely have an ejecta origin. The ionization timescales of RCW 103 are generally greater than those of G11.2$-$0.3 and Kes 73, consistent with an older age for this source. The mean ionization timescale from our fits is $ \langle n_{e}t \rangle = 9.25 \times 10^{11}$ s cm$^{-3}$, indicating the plasma in RCW 103 is approaching collisional ionization equilibrium.

\begin{deluxetable}{cccccccc}
\tablecolumns{8} \tablewidth{0pc} 
\tabletypesize{\scriptsize}
\tablecaption{Spectral Results for G11.2$-$0.3}
\tablehead{ \colhead{Region} & \colhead{$N_{H}$} & \colhead{$kT$} & \colhead{${\rm Mg}/{\rm Mg}_{\sun}$} & \colhead{${\rm Si}/{\rm Si}_{\sun}$} & \colhead{${\rm S}/{\rm S}_{\sun}$} & \colhead{$n_{e}t$} & \colhead{$\chi^{2}$/d.o.f.} \\
\colhead{} & \colhead{($\times$10$^{22}$ cm$^{-2}$)} & \colhead{(keV)} & \colhead{} & \colhead{} & \colhead{} & \colhead{(s cm$^{-3}$)} & \colhead{} }
\startdata
A & 2.3 & 0.62$\pm$0.02 & 1.30$\pm$0.12 & 1.59$\pm$0.10 & 1.82$\pm$0.14 & 6.61e11 & 655/625 \\
B & 2.4 & 0.85$\pm$0.06 & 1.04$\pm$0.21 & 1.28$\pm$0.19 & 1.29$\pm$0.22 & 1.44e11 & 429/443 \\
C & 2.1 & 0.98$\pm$0.06 & 1.05$\pm$0.14 & 1.49$\pm$0.19 & 1.03$\pm$0.17 & 1.11e11 & 579/597 \\
D & 2.7 & 1.38$\pm$0.05 & 2.80$\pm$0.58 & 3.93$\pm$0.92 & 2.51$\pm$0.57 & 1.47e11 & 462/512 \\
E & 2.6 & 0.70$\pm$0.04 & 0.87$\pm$0.16 & 1.38$\pm$0.16 & 1.85$\pm$0.21 & 3.16e11 & 551/547 \\
F & 2.7 & 0.62$\pm$0.03 & 1.13$\pm$0.21 & 1.35$\pm$0.13 & 1.40$\pm$0.16 & 9.20e11 & 633/636 \\
G & 2.3 & 0.69$\pm$0.04 & 0.90$\pm$0.15 & 1.44$\pm$0.15 & 1.60$\pm$0.19 & 4.99e11 & 559/585 \\
H & 2.6 & 0.70$\pm$0.04 & 1.07$\pm$0.23 & 1.14$\pm$0.15 & 1.23$\pm$0.17 & 7.22e11 & 500/554 \\
I & 2.4 & 0.83$\pm$0.03 & 1.28$\pm$0.21 & 1.72$\pm$0.21 & 1.69$\pm$0.18 & 3.52e11 & 691/718 \\
J & 2.5 & 0.80$\pm$0.03 & 1.34$\pm$0.22 & 1.84$\pm$0.23 & 1.71$\pm$0.19 & 4.22e11 & 469/448 \\
K & 2.4 & 0.69$\pm$0.02 & 1.23$\pm$0.14 & 1.40$\pm$0.11 & 1.79$\pm$0.14 & 9.09e11 & 658/600 \\
L & 2.5 & 1.12$\pm$0.06 & 1.74$\pm$0.37 & 2.16$\pm$0.42 & 1.74$\pm$0.29 & 1.70e11 & 501/542 \\
M & 2.7 & 0.63$\pm$0.02 & 1.19$\pm$0.21 & 1.39$\pm$0.14 & 1.39$\pm$0.15 & 1.02e12 & 664/675 \\
N & 2.7 & 0.71$\pm$0.04 & 1.68$\pm$0.52 & 2.35$\pm$0.50 & 1.79$\pm$0.36 & 7.80e11 & 361/422 \\
O & 2.9 & 0.72$\pm$0.04 & 1.21$\pm$0.29 & 1.40$\pm$0.19 & 1.48$\pm$0.19 & 3.35e11 & 534/550 \\
P & 2.7 & 0.66$\pm$0.02 & 1.08$\pm$0.11 & 1.35$\pm$0.08 & 1.21$\pm$0.09 & 7.39e11 & 1041/942 \\
Q & 2.2 & 0.64$\pm$0.01 & 1.45$\pm$0.10 & 1.58$\pm$0.08 & 1.41$\pm$0.09 & 1.03e12 & 943/819 \\
R & 2.1 & 0.60$\pm$0.01 & 1.71$\pm$0.14 & 2.03$\pm$0.12 & 1.98$\pm$0.16 & 7.24e11 & 986/892 \\
S & 2.3 & 0.83$\pm$0.03 & 1.80$\pm$0.20 & 2.30$\pm$0.23 & 2.13$\pm$0.22 & 1.72e11 & 852/769 \\
T & 2.4 & 0.64$\pm$0.02 & 2.07$\pm$0.32 & 2.19$\pm$0.26 & 2.21$\pm$0.28 & 6.34e11 & 620/627 \\
U & 2.1 & 0.79$\pm$0.03 & 1.52$\pm$0.15 & 1.47$\pm$0.14 & 1.19$\pm$0.15 & 2.30e11 & 817/761 \\
V & 2.2 & 1.37$\pm$0.09 & 2.40$\pm$0.29 & 2.40$\pm$0.32 & 1.46$\pm$0.23 & 8.90e10 & 740/712 \\
W & 2.5 & 0.72$\pm$0.04 & 0.92$\pm$0.15 & 1.05$\pm$0.12 & 1.00$\pm$0.17 & 2.20e11 & 525/529 \\
\enddata
\end{deluxetable}

\begin{deluxetable}{cccccccc}
\tablecolumns{8} \tablewidth{0pc} 
\tabletypesize{\scriptsize}
\tablecaption{Spectral Results for Kes 73}
\tablehead{ \colhead{Region} & \colhead{$N_{H}$} & \colhead{$kT$} & \colhead{${\rm Mg}/{\rm Mg}_{\sun}$} & \colhead{${\rm Si}/{\rm Si}_{\sun}$} & \colhead{${\rm S}/{\rm S}_{\sun}$} & \colhead{$n_{e}t$} & \colhead{$\chi^{2}$/d.o.f.} \\
\colhead{} & \colhead{($\times$10$^{22}$ cm$^{-2}$)} & \colhead{(keV)} & \colhead{} & \colhead{} & \colhead{} & \colhead{(s cm$^{-3}$)} & \colhead{} }
\startdata
A & 3.0 & 0.73$\pm$0.06 & 1.75$\pm$0.71 & 1.56$\pm$0.45 & 2.13$\pm$0.51 & 6.87e11 & 88/91 \\
B & 2.9 & 0.97$\pm$0.13 & 2.56$\pm$1.03 & 2.52$\pm$0.88 & 2.10$\pm$0.59 & 2.12e11 & 107/96 \\
C & 3.2 & 0.80$\pm$0.07 & 1.64$\pm$0.63 & 1.78$\pm$0.47 & 1.29$\pm$0.31 & 4.87e11 & 108/101 \\
D & 3.5 & 0.89$\pm$0.07 & 2.93$\pm$1.22 & 2.38$\pm$0.71 & 1.91$\pm$0.44 & 2.31e11 & 163/127 \\
E & 2.8 & 0.78$\pm$0.04 & 1.33$\pm$0.24 & 1.52$\pm$0.18 & 1.55$\pm$0.16 & 3.24e11 & 220/163 \\
F & 2.6 & 0.94$\pm$0.07 & 1.27$\pm$0.37 & 1.49$\pm$0.28 & 1.47$\pm$0.21 & 4.47e11 & 156/137 \\
G & 2.6 & 0.99$\pm$0.10 & 1.19$\pm$0.38 & 1.68$\pm$0.33 & 1.81$\pm$0.26 & 2.60e11 & 132/123 \\
H & 3.0 & 0.72$\pm$0.06 & 1.63$\pm$0.54 & 1.65$\pm$0.34 & 1.67$\pm$0.30 & 2.50e11 & 110/103 \\
I & 3.1 & 0.77$\pm$0.09 & 2.56$\pm$1.15 & 2.21$\pm$0.71 & 2.50$\pm$0.61 & 4.37e11 & 105/89 \\
J & 3.2 & 0.76$\pm$0.06 & 1.22$\pm$0.49 & 1.29$\pm$0.26 & 0.87$\pm$0.17 & 4.17e11 & 122/114 \\
K & 2.7 & 0.79$\pm$0.05 & 1.27$\pm$0.30 & 1.53$\pm$0.24 & 1.35$\pm$0.20 & 5.73e11 & 155/127 \\
L & 2.6 & 0.73$\pm$0.04 & 1.31$\pm$0.33 & 1.24$\pm$0.20 & 1.30$\pm$0.21 & 1.64e12 & 154/127 \\
M & 2.2 & 0.85$\pm$0.05 & 1.05$\pm$0.20 & 1.43$\pm$0.21 & 1.85$\pm$0.23 & 5.12e11 & 136/135 \\
N & 2.5 & 1.12$\pm$0.20 & 1.07$\pm$0.47 & 1.74$\pm$0.51 & 2.67$\pm$0.79 & 7.75e10 & 88/62 \\
O & 2.7 & 0.75$\pm$0.07 & 0.56$\pm$0.17 & 0.99$\pm$0.16 & 1.65$\pm$0.34 & 1.27e11 & 91/84 \\
P & 3.4 & 0.66$\pm$0.07 & 2.24$\pm$1.48 & 1.41$\pm$0.51 & 1.87$\pm$0.54 & 5.24e12 & 92/81 \\
Q & 2.5 & 0.70$\pm$0.07 & 0.86$\pm$0.30 & 1.25$\pm$0.27 & 1.57$\pm$0.35 & 5.04e11 & 80/80 \\
R & 2.8 & 0.79$\pm$0.07 & 0.48$\pm$0.20 & 1.03$\pm$0.25 & 1.30$\pm$0.29 & 4.16e11 & 102/86 \\
S & 3.0 & 0.70$\pm$0.08 & 1.34$\pm$0.64 & 1.23$\pm$0.33 & 1.26$\pm$0.32 & 3.39e11 & 57/74 \\
T & 2.8 & 0.89$\pm$0.14 & 1.44$\pm$0.69 & 1.33$\pm$0.45 & 1.81$\pm$0.48 & 3.09e11 & 59/67 \\
U & 2.7 & 1.12$\pm$0.14 & 1.64$\pm$0.46 & 2.05$\pm$0.57 & 1.73$\pm$0.44 & 1.87e11 & 113/103 \\
V & 2.5 & 0.63$\pm$0.03 & 1.50$\pm$0.37 & 1.50$\pm$0.22 & 1.41$\pm$0.25 & 3.20e12 & 136/123 \\
W & 2.3 & 0.86$\pm$0.10 & 0.99$\pm$0.31 & 1.64$\pm$0.32 & 1.45$\pm$0.28 & 4.14e11 & 104/96 \\
X & 2.6 & 0.70$\pm$0.09 & 1.53$\pm$0.56 & 2.07$\pm$0.45 & 1.51$\pm$0.35 & 5.98e11 & 118/85 \\
Y & 2.8 & 0.89$\pm$0.17 & 0.87$\pm$0.50 & 1.98$\pm$0.65 & 1.38$\pm$0.42 & 3.10e11 & 53/56 \\
Z & 2.3 & 0.65$\pm$0.05 & 1.23$\pm$0.35 & 1.54$\pm$0.25 & 1.99$\pm$0.34 & 4.50e12 & 122/110 \\
a & 2.3 & 0.88$\pm$0.08 & 0.87$\pm$0.19 & 1.07$\pm$0.19 & 1.00$\pm$0.21 & 2.14e11 & 123/106 \\
b & 2.1 & 0.89$\pm$0.11 & 0.98$\pm$0.24 & 1.18$\pm$0.22 & 1.12$\pm$0.26 & 2.10e11 & 98/87 \\
c & 2.3 & 0.84$\pm$0.10 & 0.92$\pm$0.28 & 1.28$\pm$0.32 & 1.09$\pm$0.30 & 3.28e11 & 80/79 \\
d & 2.2 & 1.22$\pm$0.16 & 1.13$\pm$0.33 & 1.11$\pm$0.24 & 1.37$\pm$0.31 & 1.09e11 & 110/95 \\
\enddata
\end{deluxetable}

\begin{deluxetable}{cccccccc}
\tablecolumns{8} \tablewidth{0pc}
\tabletypesize{\scriptsize}
\tablecaption{Spectral Results for RCW 103}
\tablehead{ \colhead{Region} & \colhead{$N_{H}$} & \colhead{$kT$} & \colhead{${\rm Fe}/{\rm Fe}_{\sun}$} & \colhead{${\rm Mg}/{\rm Mg}_{\sun}$} & \colhead{${\rm Si}/{\rm Si}_{\sun}$} & \colhead{$n_{e}t$} & \colhead{$\chi^{2}$/d.o.f.} \\
\colhead{} & \colhead{($\times$10$^{22}$ cm$^{-2}$)} & \colhead{(keV)} & \colhead{} & \colhead{} & \colhead{} & \colhead{(s cm$^{-3}$)} & \colhead{} }
\startdata
A & 0.6 & 0.52$\pm$0.03 & 1.35$\pm$0.23 & 1.75$\pm$0.26 & 2.19$\pm$0.37 & 5.61e11 & 60/63 \\
B & 0.8 & 0.53$\pm$0.03 & 0.98$\pm$0.15 & 1.19$\pm$0.17 & 1.69$\pm$0.25 & 6.43e11 & 74/88 \\
C & 0.6 & 0.57$\pm$0.01 & 1.05$\pm$0.13 & 1.47$\pm$0.16 & 1.68$\pm$0.21 & 7.49e11 & 116/79 \\
D & 0.5 & 0.55$\pm$0.01 & 0.97$\pm$0.10 & 1.42$\pm$0.13 & 1.38$\pm$0.17 & 1.88e12 & 98/88 \\
E & 0.6 & 0.52$\pm$0.01 & 0.88$\pm$0.08 & 1.41$\pm$0.11 & 1.56$\pm$0.16 & 1.88e12 & 116/91 \\
F & 0.7 & 0.52$\pm$0.03 & 1.65$\pm$0.20 & 1.84$\pm$0.22 & 1.84$\pm$0.26 & 2.70e11 & 126/92 \\
G & 0.7 & 0.61$\pm$0.03 & 1.09$\pm$0.17 & 1.70$\pm$0.23 & 1.12$\pm$0.21 & 5.17e11 & 122/86 \\
H & 0.7 & 0.67$\pm$0.03 & 1.52$\pm$0.29 & 1.86$\pm$0.28 & 1.40$\pm$0.27 & 2.67e11 & 126/91 \\
I & 0.6 & 0.53$\pm$0.02 & 1.28$\pm$0.16 & 1.51$\pm$0.18 & 1.82$\pm$0.25 & 4.73e11 & 102/88 \\
J & 0.8 & 0.60$\pm$0.02 & 1.09$\pm$0.19 & 1.84$\pm$0.27 & 1.66$\pm$0.28 & 7.24e11 & 105/88 \\
K & 0.7 & 0.57$\pm$0.02 & 1.11$\pm$0.14 & 1.40$\pm$0.16 & 1.12$\pm$0.17 & 5.76e11 & 86/76 \\
L & 1.0 & 0.33$\pm$0.06 & 1.58$\pm$0.44 & 1.51$\pm$0.31 & 2.94$\pm$1.20 & 6.83e11 & 72/68 \\
M & 0.5 & 0.55$\pm$0.02 & 0.92$\pm$0.26 & 1.72$\pm$0.37 & 1.40$\pm$0.43 & 2.04e12 & 122/82 \\
N & 0.6 & 0.59$\pm$0.02 & 0.91$\pm$0.12 & 1.31$\pm$0.15 & 1.41$\pm$0.19 & 6.73e11 & 114/100 \\
O & 0.5 & 0.56$\pm$0.01 & 0.81$\pm$0.03 & 1.12$\pm$0.04 & 1.28$\pm$0.06 & 1.07e12 & 348/126 \\
P & 0.5 & 0.58$\pm$0.02 & 0.62$\pm$0.06 & 1.02$\pm$0.10 & 1.03$\pm$0.12 & 1.54e12 & 112/108 \\
Q & 0.4 & 0.55$\pm$0.01 & 0.67$\pm$0.05 & 1.08$\pm$0.08 & 1.13$\pm$0.11 & 1.57e12 & 145/98 \\
R & 0.5 & 0.57$\pm$0.01 & 0.89$\pm$0.08 & 1.15$\pm$0.10 & 1.08$\pm$0.11 & 7.70e11 & 148/110 \\
S & 0.5 & 0.53$\pm$0.02 & 1.24$\pm$0.14 & 1.43$\pm$0.15 & 1.82$\pm$0.22 & 7.57e11 & 111/99 \\
T & 0.5 & 0.60$\pm$0.01 & 0.83$\pm$0.08 & 1.06$\pm$0.10 & 1.05$\pm$0.13 & 1.09e12 & 126/91 \\
U & 0.3 & 0.63$\pm$0.03 & 0.56$\pm$0.24 & 1.47$\pm$0.46 & 1.83$\pm$0.68 & 1.71e12 & 53/69 \\
V & 0.3 & 0.58$\pm$0.01 & 0.73$\pm$0.07 & 1.03$\pm$0.09 & 1.11$\pm$0.13 & 9.08e11 & 147/105 \\
W & 0.4 & 0.55$\pm$0.01 & 0.76$\pm$0.06 & 1.00$\pm$0.08 & 0.98$\pm$0.11 & 1.03e12 & 131/107 \\
X & 0.4 & 0.59$\pm$0.01 & 1.00$\pm$0.09 & 1.29$\pm$0.11 & 1.42$\pm$0.15 & 4.19e11 & 153/101 \\
Y & 0.4 & 0.54$\pm$0.02 & 0.91$\pm$0.09 & 1.14$\pm$0.11 & 1.21$\pm$0.16 & 4.69e11 & 88/96 \\
Z & 0.4 & 0.53$\pm$0.01 & 1.02$\pm$0.10 & 1.34$\pm$0.12 & 1.18$\pm$0.15 & 6.58e11 & 133/100 \\
a & 0.5 & 0.57$\pm$0.01 & 0.85$\pm$0.07 & 1.18$\pm$0.09 & 1.45$\pm$0.13 & 7.95e11 & 127/97 \\
b & 0.4 & 0.59$\pm$0.02 & 1.20$\pm$0.13 & 1.50$\pm$0.15 & 1.41$\pm$0.18 & 4.68e11 & 127/84 \\
c & 0.5 & 0.57$\pm$0.02 & 0.55$\pm$0.08 & 1.23$\pm$0.14 & 1.21$\pm$0.18 & 2.06e12 & 80/77 \\
d & 0.5 & 0.66$\pm$0.01 & 1.04$\pm$0.10 & 1.75$\pm$0.15 & 1.52$\pm$0.16 & 5.34e11 & 115/93 \\
e & 0.4 & 0.54$\pm$0.01 & 1.11$\pm$0.11 & 1.75$\pm$0.16 & 1.78$\pm$0.21 & 9.05e11 & 86/87 \\
\enddata
\end{deluxetable} 

\end{document}